\documentclass[apj]{emulateapj}

\shorttitle{Twin Binaries}
\shortauthors{Lombardi et al.}

\begin{document}
\title{Twin Binaries: Studies of Stability, Mass Transfer, and Coalescence}
\author{J.\ C.\ Lombardi, Jr.$^1$,
W.\ Holtzman$^1$,
K.\ L.\ Dooley$^2$,
K.\ Gearity$^1$,
V.\ Kalogera$^{3,4}$,
F.\ A.\ Rasio$^{3,4}$}
\altaffiltext{1}{Department of Physics, Allegheny
College, Meadville, PA, 16335}
\altaffiltext{2}{Department of Physics, University of Florida, Gainesville, FL 32611}
\altaffiltext{3}{Department of Physics and
Astronomy, Northwestern University, 2145 Sheridan Rd., Evanston, IL 60208}
\altaffiltext{4}{Center for Interdisciplinary Exploration and Research in Astrophysics (CIERA), Northwestern University.}

\begin{abstract}
Motivated by suggestions that binaries with almost equal-mass
components (``twins") play an important role in the formation of
double neutron stars and may be rather abundant among binaries, we
study the stability of synchronized close and contact binaries with
identical components in circular orbits.  In particular, we
investigate the dependency of the innermost stable circular orbit on
the core mass, and we study the coalescence of the binary that occurs
at smaller separations.  For twin binaries composed of convective
main-sequence stars, subgiants, or giants with low mass cores
($M_c\lesssim 0.15M$, where $M$ is the mass of a component), a secular
instability is reached during the contact phase, accompanied by a
dynamical mass transfer instability at the same or at a slightly
smaller orbital separation.  Binaries that come inside this
instability limit transfer mass gradually from one component to the
other and then coalesce quickly as mass is lost through the outer
Lagrangian points.  For twin giant binaries with moderate to massive
cores ($M_c\gtrsim 0.15M$), we find that stable contact configurations
exist at all separations down to the Roche limit, when mass shedding
through the outer Lagrangian points triggers a coalescence of the
envelopes and leaves the cores orbiting in a central tight binary.  In
addition to the formation of binary neutron stars, we also discuss the
implications of our results for the production of planetary nebulae
with double degenerate central binaries.
\end{abstract}


\keywords{binaries: close --- binaries: general --- hydrodynamics --- instabilities --- methods: numerical --- stars: general}

\section{Introduction and Motivation}

\subsection{Formation of Binary Neutron Stars}

The evolutionary history and formation of close binaries with two
neutron stars similar to the Hulse-Taylor pulsar B1913+16
\citep{1975ApJ...195L..51H} and the double pulsar J0737-3039
\citep{2003Natur.426..531B} is a topic of intense current
interest. Most recent studies of the known double neutron stars focus
on the stages going back to the time of the second supernova explosion
and the formation of the youngest of the two neutron stars
\citep[][and references
therein]{2004MNRAS.349..169D,2004ApJ...603L.101W,2004ApJ...616..414W,2005PhRvL..94e1102P,2006MNRAS.373L..50S,2006ApJ...639.1007W,2010ApJ...721.1689W}. Although
these studies provide very interesting constraints on the properties
of the stellar progenitor of the second neutron star, they do not
probe the earlier evolutionary history. That part remains uncertain
and more difficult to constrain empirically based on the measured
properties of observed systems.

Since the discovery of the Hulse-Taylor binary, the origin of double
neutron stars has been naturally connected to the evolution of massive
binaries, with stellar components that are massive enough to form two
neutron stars at the end of their lifetime. Over the years, a
qualitative consensus of understanding for the formation of double
neutron stars developed \citep[see, e.g.,][]{1991PhR...203....1B}:
massive binaries experience a phase of stable mass transfer when the
primary overflows its Roche lobe revealing its helium core; this core ends
its lifetime in a supernova forming the first neutron star in the
system; the binary becomes a high-mass X-ray binary until the massive
secondary fills its Roche lobe and the binary enters a dynamically
unstable phase of mass transfer leading to inspiral in a common
envelope that engulfs the neutron star; during this phase the neutron
star is thought to be spun up through recycling and, if the binary
avoids a merger in the inspiral, the helium core of the secondary is
revealed after the envelope ejection; this core explodes in a
supernova, forming the second neutron star in the system, and the double
neutron star further evolves through orbital contraction and
gravitational-wave emission. Variations of this evolutionary sequence
have been shown to be realized by theoretical binary population
studies \citep[e.g.,][]{2002ApJ...572..407B}.

\cite{1995ApJ...440..270B} 
argued that the inspiral of the neutron
star during the common envelope phase in the standard model is
problematic: the neutron star is expected to experience hypercritical
accretion \citep{1993ApJ...411L..33C} at rates many orders of
magnitude above the photon Eddington limit through a neutrino-cooled
accretion flow \citep[see also][]{1996ApJ...460..801F}. Such a rapid
accretion phase along with the adoption of a low maximum neutron star
mass ($\sim$1.5\,M$_\odot$ derived for a soft equation of state for
neutron star matter with kaon condensation) led Brown to conclude
that all neutron stars in common envelope phases will accrete enough
matter to collapse into a black hole. Consequently, he argued that
the ``standard'' evolutionary sequence described above aborts the
formation of double neutron stars and instead leads to the formation
of binaries with low-mass black holes and neutron star companions.
We note however that, even with the same treatment of hypercritical accretion,
a maximum neutron star mass of $\sim2 M_\odot$
(corresponding to a more regular stiff equation of state) does prevent a good
fraction of neutron stars from being transformed into low-mass
black holes \citep{2002ApJ...572..407B}.

\cite{1995ApJ...440..270B} also noted how the masses
measured in known double neutron stars are very close to being equal
\citep[][and references
therein]{1996ApJ...466L..87N,1999ApJ...512..288T,2002ApJ...581..501S,2005ASPC..328...25W,2006ApJ...644L.113J,2006Sci...314...97K}.
Motivated by these two points, he proposed a different formation
channel for double neutron stars.  Brown suggested that
double neutron stars form from massive binaries with component masses
that are within $\sim$4\% of one another.  Consequently the red giant
phases of the two components coincide in time and when mass transfer
ensues from the primary, both components have deep convective
envelopes and well developed helium cores, so that a {\em double} core
phase develops, where the two helium cores orbit within the combined
envelopes of the two massive stars. Provided that there is enough
orbital energy, the common envelope is ejected before the two cores
merge, and a tight binary with two helium cores is formed. These two
cores differ very little in mass and reach core collapse one very soon
after the other \citep[$\sim 10^3$\,yr based on helium-star
models,][]{1986A&A...167...61H,1994A&A...290..119P}, forming a close
double neutron star.

The advantages of this hypothesized evolutionary
channel are (i) a neutron star never experiences common envelope
spiral-in and hypercritical accretion, and (ii) the two stellar
components have so similar masses that they naturally form neutron
stars of almost equal mass as observed
\citep{1998ApJ...506..780B,2007PhR...442....5B}. On the other hand,
this channel (i) requires that mass transfer between the red giant
progenitors will indeed lead to the inspiral of the two cores in a
common envelope and (ii) requires fine-tuning the conditions for
recycling the first neutron star, as this spin up must occur during
the very short interval between the two supernovae through the stellar
wind or possibly during the brief Roche-lobe overflow from the
lower-mass helium star \citep{2006MNRAS.368.1742D}.  A potential additional
disadvantage is that this channel is very restrictive in that it
requires progenitors that are at most only $\sim 4$\% apart in mass: however,
in their study of protobinary stars, \cite{2007ApJ...661.1034K} 
find that for a wide range of initial conditions Roche
lobe overflow tends to equalize the masses of the binary components.

The double neutron star formation channel suggested by
\cite{1995ApJ...440..270B} has attracted renewed attention because of
the reported abundance of ``twins,'' massive binaries with mass ratios
very close to unity (within 5\%) by
\cite{2006ApJ...639L..67P}. Specifically they analyze data for 21
detached eclipsing binaries in the Small Magellanic Cloud and find
that the data are consistent with a flat mass function containing 55\%
of the systems and ``twins'' with mass ratios greater than 0.95
containing the remaining 45\% of the population. However, it is
important to note that there are severe selection effects against the
discovery of binaries with small mass ratios \citep[typically
$\lesssim
0.5$,][]{1992Ap&SS.194..143H,1992Ap&SS.195..359H,1992Ap&SS.196..299H};
therefore the contribution of twins may not be as significant as
implied by the most recent observations. Quantitative modeling of the
associated selection effects is required to derive more reliable
statistical conclusions.

Apart from uncertainties with the initial properties of the binary
population, a number of physical processes related to these formation
channels make it hard to assess their relative contribution to double
neutron star formation. The physics of common envelopes and
hypercritical accretion, as well as of neutron star equations of state
and their maximum mass, is not well understood \citep[but
see][]{2007ApJ...670..741L}.  Also, the development of a dynamical
instability and subsequent common envelope phase with the inspiral of
the two cores is assumed, but has not been investigated before in any
detail.  In this study, we attempt to understand better one of the
aspects related to the formation channel suggested by
\cite{1995ApJ...440..270B}: the fate of mass transfer between two stars
of almost equal mass.

\subsection{Binaries and Planetary Nebulae}

Although our primary motivation is to investigate a formation channel
for binary neutron stars, our results are relevant to other scenarios
as well.  For example, through common envelope evolution, binaries could transform quite
naturally into planetary nebulae (PNe) with a central close
binary.  The influence of binaries on the production and morphology of
PNe has received increased attention in recent years
\citep[e.g.,][]{2009A&A...505..249M,2010MNRAS.408.2312J},
and observations to date have identified approximately 40 close binaries in the
centers of PNe: see \citet{2009PASP..121..316D} for a summary
of both the relevant theory and observations.
As lifetimes of PNe are less than
only $10^5$ years, an embedded central binary has undergone no significant evolution since the 
common envelope phase that presumably formed it.

About a quarter of the known close binary systems within PNe are
thought to be double degenerates, that is, binaries in which the
components are pre-white dwarfs (also known as extreme horizontal
branch stars, hot subdwarfs, or subdwarf B and O stars) or white
dwarfs.
Such double degenerate binaries, with or without planetary
nebulae, are particularly interesting.  As possible progenitors for
Type Ia supernovae, they are the desired targets of the ESO SN Ia
Progenitor Survey \citep[see][and references
  therein]{2011A&A...528L..16G}.

Interestingly, three of the five well studied double degenerate binaries within PNe \citep[see][and references therein]{2011apn5.confE.275H}
may contain nearly equal mass components and therefore
are possibly the progeny of twin binaries.
In particular, models of a double hot
subdwarf binary with $q\approx 1$ are consistent with
photometric and spectroscopic observations of the central stars in NGC
6026 \citep{2010AJ....140..319H}
and in Abell 41 \citep{2001A&A...377..898B,2008ARep...52..479S}.  Similarly, recent observations of the
central star in Hen 2-428 reveal it also to be a double degenerate binary,
with the effective temperature of the components being within a few
thousand Kelvin \citep{2011apn5.confE.259S}, suggesting a mass ratio
near $q=1$.
  As we investigate in this paper, the coalescence
of a twin giant binary could lead
to the formation of double degenerate core surrounded by a circumbinary envelope,
a type of proto-PN.

\subsection{Theoretical Work on Close Binaries}

Most of the classical theoretical work on close binaries was done in the limit of
a self-gravitating incompressible fluid \citep[see][and references
therein]{1969efe..book.....C}.  An essential result found in the
incompressible case is that the hydrostatic equilibrium solutions for
sufficiently close binaries can become globally unstable
\citep{1975ApJ...202..809C,1975ApJ...202..803T}.  The classical
analytic studies for binaries were extended to polytropes in the work
of
\citet{1993ApJS...88..205L,1993ApJ...406L..63L,1994ApJ...420..811L,1994ApJ...423..344L,1994ApJ...437..742L}.
In their approach, the stars are modeled as self-gravitating
compressible ellipsoids, and an energy variational principle is used
to construct approximate equilibrium configurations and study their
stability.  These treatments, along with complementary numerical
hydrodynamic calculations
\citep{1992ApJ...401..226R,1994ApJ...432..242R,1995ApJ...438..887R},
demonstrate that dynamical instabilities persist in the compressible
regime and can cause a binary to coalesce to form a rapidly rotating
spheroidal object.  Such a dynamical instability can trigger a merger
in just a few orbital periods \citep{1992ApJ...401..226R} or an
episode of mass transfer that lasts many orbits
\citep{2002ApJS..138..121M,2006ApJ...643..381D,2008NewAR..51..878F,2009JPhCS.172a2034D,2009A&A...500.1193L}.


The evolution of a close binary system can also be affected by another
type of global instability.  It has been referred to by various names,
such as the secular instability
\citep{1993ApJS...88..205L,1993ApJ...406L..63L,1994ApJ...420..811L,1994ApJ...423..344L,1994ApJ...437..742L},
tidal instability \citep{1973ApJ...180..307C,1980A&A....92..167H},
gravogyro instability \citep{1984PASJ...36..239H}, and Darwin
instability \citep{1993ApJ...410..328L}.  Its physical origin is easy
to understand
\citep{1993ApJS...88..205L,1994ApJ...423..344L,1994MmSAI..65...37R,2006JAVSO..35..124W}. There
exists a minimum value of the total angular momentum $J$ for a
synchronized close binary. This is simply because the spin angular
momentum, which increases as the separation $r$ decreases for a
synchronized system, can become comparable to the orbital angular
momentum for sufficiently small $r$. A system that reaches the minimum
of $J$ cannot evolve further by angular momentum loss and remain
synchronized. Instead, the combined action of tidal forces and viscous
dissipation will drive the system out of synchronization and cause
rapid orbital decay as angular momentum is continually transferred
from the orbit to the spins.  The orbital decay then proceeds on a
timescale comparable to the synchronization time of a stable binary.

In this paper, we pay particular attention to the onset of orbital
instabilities, including those characterized by mass transfer, as well as
the subsequent inspiral of the cores.  We extend previous hydrodynamic
studies of close binary systems to cases that involve identical
giants, that is, to twin stars with dense stellar cores and extended
envelopes.  So that our results can be scaled to stars of any size or
mass, we approximate each star as a condensed polytrope, namely, as a
point mass surrounded by a uniform specific entropy fluid of adiabatic index
$\Gamma=5/3$.  Such a model is appropriate for a fully convective
monatomic ideal gas surrounding a compact core.  We consider
fractional core masses $m_c$ that cover the entire range of
theoretical possibilities.  Zero core mass models correspond to
normal (or ``complete'') $n=1.5$ polytropes, which are most appropriate
for low mass main-sequence stars and non-relativistic white
dwarfs.  At the other extreme,
when $m_c=1$, there is no mass in the gaseous envelope, and the binary
is simply that of two point masses.  By varying the core mass between
these extremes, we are able to study in a systematic way the full
parameter space of twin binaries and determine under what conditions
mass transfer develops or an innermost stable circular orbit exists.

For typical compositions, the subgiant phase begins when the core mass
grows to $\sim 10$\% of the total mass, the so-called Sch{\"
o}nberg-Chandrasekhar limit \citep{1942ApJ....96..161S}.  By the time
the star reaches the base of the red giant branch, the core mass has
increased by a few more percent of the total mass.  Roughly speaking
then, our models with $m_c\lesssim 0.1$, $0.1 \lesssim m_c \lesssim
0.13$, and $m_c\gtrsim 0.13$ correspond to main sequence, subgiant,
and giant stars, respectively.  The assumption that the stellar
envelope has constant specific entropy makes our models most relevant
to stars that are fully convective (as with low mass main sequence
stars) or that have deep convective envelopes (as in many red giants).

The use of condensed polytropes as a model of red giants has a rich
history, including seminal work by \cite{1939isss.book.....C},
\cite{1953ApJ...118..529O}, and \cite{1955ApJS....1..319H}.  Mass
transfer in close binary systems has been modeled with the help of
condensed polytropes as well: the response of the donor due to mass
loss is considered, with its resulting contraction, or expansion,
compared against that of its Roche lobe
\citep{1972AcA....22...73P,1987ApJ...318..794H}.  If at the onset of
mass transfer the star contracts less rapidly than its Roche lobe (or
expands more rapidly than it), then the mass exchange is dynamically
unstable, and the binary will evolve rapidly toward a new, often
qualitatively different, equilibrium.  \cite{1987ApJ...318..794H} find
that equal mass condensed polytrope contact binaries with fractional
core masses $m_c\lesssim 0.46$ experience stable mass transfer.
More recently, \cite{2007ApJ...661.1034K} have used condensed
polytropes to study the formation of twin star systems.

Although essential for a qualitative understanding of mass transfer,
such treatments of close binaries do, however, make several
simplifying approximations: most importantly, (i) the dynamics of the
orbit and size of the Roche lobe are treated in the point mass
approximation, (ii) the response of the binary components to mass loss
or gain is modeled as if each star were spherical and in isolation,
and (iii) mass that overflows a Roche lobe is considered to leave that
star.  These approximations are quite reasonable for semidetached
binaries, but their validity can be questioned for contact binaries.
In such cases, a common envelope persists in equilibrium outside of
the Roche lobes (the inner Lagrangian surface) so that the pressure
and density on the Roche lobes are non-zero: mass that overflows a
Roche lobe is {\it not} necessarily transferred to the other star but
rather can persist in equilibrium inside the outer Lagrangian surface.
A primary goal of this paper is therefore to relax the approximations
of previous works by using accurate hydrodynamical calculations to
study contact binary systems.

Our paper is organized as follows.  In \S2 we review our numerical
method and general conventions.  In \S3 we present our results for the
equilibrium and stability properties of twin binary systems.
The dynamical evolution to complete coalescence is
followed for several unstable systems.  Implications of our
results are discussed in \S4.

\section{Numerical Methods and Assumptions\label{num_methods}}

\subsection{The SPH code}

To generate our models, we use a modified version of the SPH code
originally developed by \citet{ras91} that has been updated to include
the variational equations of motion derived in
\cite{2009arXiv0904.0997G}.  SPH is a Lagrangian particle method,
meaning that the fluid is represented by a finite number of fluid
elements or ``particles.''  Associated with each particle $i$ are, for
example, its position ${\bf r}_i$, velocity ${\bf v}_i$, and mass
$m_i$.  Each particle also carries a purely numerical smoothing length
$h_i$ that determines the local spatial resolution and is used in the
calculation of fluid properties such as acceleration and density.
Details of our SPH code, such as the particular form of the artificial
viscosity $\Pi_{ij}$ and smoothing kernel $W_{ij}$ implemented, are
described in \cite{2009arXiv0904.0997G}.  See \citet{ras99} and
\citet{2009NewAR..53...78R} for reviews of SPH.

Because the gas in our initial stellar models is of constant specific entropy, we find
it convenient to integrate the so-called entropic variable $A_i$ of
each particle $i$.  The entropic variable is simply the
proportionality constant in the polytropic equation of state
$p=A\rho^\Gamma$, where $p$ is pressure and $\rho$ is density.  The
entropic variable is so named because of its close connection to
entropy: both quantities are conserved in reversible processes and
strictly increase otherwise.  We therefore use $dA_i/dt=0$ in the
relaxations of our single star models and in the calculations of our
binary equilibrium sequences.  For our dynamical calculations of
merger scenarios (see \S\ref{dynamical} and
\S\ref{dynamical_integrations}), we evolve $A_i$ according to the
discretized SPH version of the first law of thermodynamics:
\begin{equation}
{dA_i\over dt}= {\Gamma-1\over 2\rho_i^{\Gamma-1}}
\sum_jm_j
\Pi_{ij}\,
({\bf v}_i-{\bf v}_j)\cdot{\bf\nabla}_iW_{ij}(h_i)\,.
\label{adot}
\end{equation}

To calculate the gravitational accelerations and potentials, we use
direct summation on NVIDIA graphics cards, softening with the usual
SPH kernel as in \citet{1989ApJS...70..419H}.  The use of such a
softening with finite extent (as opposed, for example, to Plummer
softening) increases the accuracy and stability of our SPH models,
consistent with the studies of \citet{2000MNRAS.314..475A} and
\citet{2001MNRAS.324..273D}.  The gravity of core points in our models
is similarly softened, applying a constant smoothing length comparable
to the minimum smoothing length in the system.

\subsection{Single Star Models\label{init_data}}

In this section we present our procedure for modeling the stars that
are used in the binary simulations of \S\ref{numerical_results}.
Hydrodynamically, a subgiant or giant can be treated as a
two-component system: a high-density, degenerate core, surrounded by
an extended envelope.  The very large density contrast between the
core and the envelope, along with the small core radius, justifies the
use of a single point mass to represent the core.  Because the giants
we wish to model mostly have deep convective envelopes with an equation of
state dominated by monatomic ideal gas pressure, we treat their gas as
a constant specific entropy fluid with an adiabatic index $\Gamma=5/3$.

Specifically, our stellar models are the so-called condensed
polytropes, namely, constant specific entropy fluid surrounding a point mass
core \citep{1939isss.book.....C,1955ApJS....1..319H}, which we
parameterize by the core mass $m_c$.  Each of the condensed polytropes
in our family of models has total mass $M=1$ and radius $R=1$.  The
unit system is completed by choosing Newton's gravitational constant
$G=1$.  Condensed polytrope models have not only the advantage of
reproducibility but also of scalability to any stellar mass and
radius: although our focus here is on stars massive enough ultimately
to yield a neutron star, the same calculations are also valid for low
mass systems.  In the limiting scenarios of $m_c=0$ and $m_c=1$, we
recover the well-studied cases of a $n=3/2$ polytrope and a point
mass, respectively.

\begin{figure}
\includegraphics[width=3.4in]{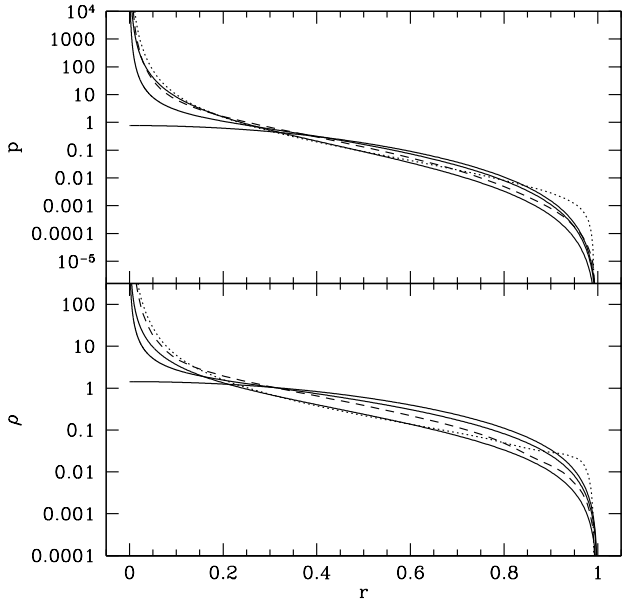}
\caption{The solid curves show the quasi-analytic radial profiles of
the pressure $p$ and density $\rho$ for $m_c=0$, 0.125, and 0.5
condensed polytropes; for comparison, the dashed and dotted curves
represent red giants, as modeled by the TWIN stellar evolution code,
with initial masses respectively of 10 and 25 $M_\odot$.  For the
condensed polytropes, curves associated with larger $m_c$ are higher
on the left edge of the figure and lower on the right edge.  Units are
such that $G=M=R=1$.
\label{rpdm}
}
\end{figure}

Figure~\ref{rpdm} shows the pressure and density profiles as a
function of radius for condensed polytropes with core masses 0, 0.125,
and 0.5.  For comparison, we also display the profiles of red giant
stars computed using the TWIN stellar evolution code
\citep{1971MNRAS.151..351E,2008A&A...488.1007G} from the MUSE software
environment\footnote{{\tt http://muse.li}}
\citep{2009NewA...14..369P}: we evolve 10 and 25 $M_\odot$ stars with
initial helium abundance $Y=0.28$ and metallicity $Z=0.02$ until they
obtain core masses of approximately $m_c=0.16$ and 0.27, respectively.
This corresponds to our initially 10 $M_\odot$ star being at the very tip of
the red giant branch and the initially 25 $M_\odot$ star being on the upper part of
the branch.  The profiles of the simple condensed polytrope models
follow the same general trend as those of the red giants, indicating
that our simple models can indeed help provide an understanding of
real red giant binaries.

Table~\ref{parents} presents the models used in this paper.  Column
(1) gives the core mass $m_c$, while column (2) lists the corresponding
$E_{\rm O}$ value: this is the parameter $E$ in the original notation of
\cite{1953ApJ...118..529O}, as well as in the notation of
\cite{1955ApJS....1..319H} and \cite{1987ApJ...318..794H}.  (We prefer
to reserve the variable $E$ for energy.)  The value of $E_{\rm
  O}$ controls the shape of the density profile: to lowest order near
the surface ($r\approx R$), $\rho(r) \approx (2/5)^{3/2}
E_{\rm O}(1-r/R)^{3/2}M/(4\pi R^3)$.  In practice, we adjust
$E_{\rm O}$ to achieve the desired core mass $m_c$.

\begin{table}[ht!]
\caption{Parent Star Characteristics}\vskip2pt
\begin{tabular}{ccc} \tableline\tableline
$m_{\rm c}$ &
$E_O$ \\
\quad(1) & (2)\\
\tableline
0     & 45.4808\\
0.05  & 39.9250\\
0.1   & 35.2403\\
0.125 & 33.1661\\
0.15  & 31.2407\\
0.175 & 29.4461\\
0.2   & 27.7673\\
0.25  & 24.7072\\
0.3   & 21.9782\\
0.4   & 17.2861\\
0.5   & 13.3582\\
0.6   &  9.9904\\
0.7   &  7.0496\\
0.8   &  4.4443\\
0.9   &  2.1092\\
0.99  &  0.1984\\
\tableline
\end{tabular}
\label{parents}
\end{table}

We begin by making an SPH model of a single star in isolation.  Unless
stated otherwise, we use $N=19938$ SPH particles initially placed on a
hexagonal close packed lattice with a lattice spacing constant
$a_1=0.0542$, with particles extending out to a radius that is between
one and two smoothing lengths less than the full stellar radius.  We
model the stellar core as a point mass that interacts gravitationally,
but not hydrodynamically, with the rest of the system, as suggested by
\cite{rs91} and others.  The gravitational influence of these core
points are softened according to the SPH kernel with $h=0.0498$.
Particle masses are first apportioned according to the desired density
profile and then slightly rescaled before relaxation begins to ensure
that the correct total mass $M=1$ is precisely achieved.  The entropic
variable $A_i$ of each SPH particle is set to the desired polytropic
constant $K$, which is determined from the mass-radius relation for
$n=3/2$ condensed polytropes \citep{1953ApJ...118..529O}:
\begin{equation}
R=(4\pi)^{-2/3}G^{-1}KE_{\rm O}^{2/3}M^{-1/3}. \label{mrrelation}
\end{equation}

After the initial parameters of the particles have been assigned, we
relax the SPH fluid into hydrostatic equilibrium.  This relaxation is
effected by including an artificial viscosity contribution in the
acceleration equation 
\citep[with $\alpha=1$ and $\beta=2$ in equation (A19) of][]{2009arXiv0904.0997G}, while still
keeping $A_i$ constant for all particles.  In this way, the entropy of
the system is preserved while the system approaches equilibrium.  The
total energy typically decreases by less than a percent in the
process, indicating that our initial assignment of particle properties
was indeed very close to an equilibrium state.

\begin{figure}
\includegraphics[width=3.4in]{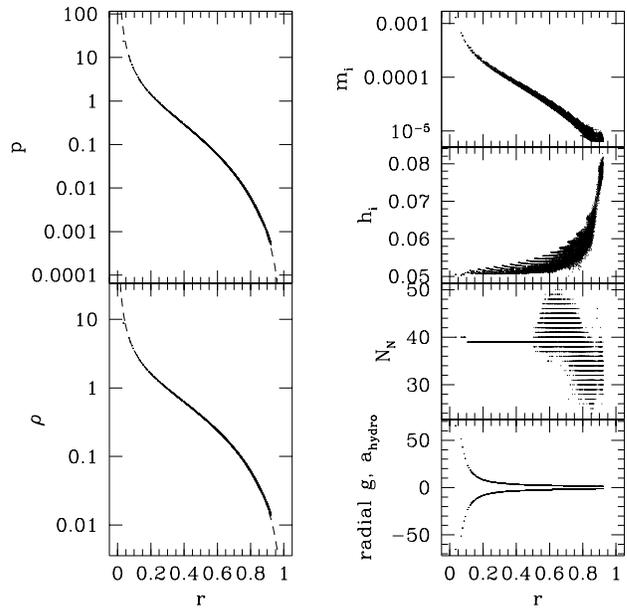}
\caption{Properties of the model with core mass $m_c=0.25$ as a
function of radius, after relaxation for 500 time units.  The frames
in the left column compare calculated pressure $p$ and density $\rho$
profiles of the star (dashed curves) against particle data from our
SPH model (dots).  The right column provides additional SPH particle
data: individual SPH particle mass $m$, smoothing length $h$, number
of neighbors $N_N$, and radial component of the hydrodynamic
acceleration $a_{\rm hydro}$ (upper data) and gravitational acceleration
$g$ (lower data).
\label{p_bw_050_mc0.25}
}
\end{figure}

This approach allows us to model the desired profiles very
accurately. An example is presented in Figure \ref{p_bw_050_mc0.25},
where we plot desired profiles and SPH particle data for our relaxed
$m_c=0.25$ star.  Although the core and surface of the star of course
cannot be resolved on the length scale of a smoothing length
(typically 0.05 to 0.08 length units), the thermodynamic profiles of
the SPH model nicely reproduce the quasi-analytic curves.  Indeed, the
SPH data in the left column of Figure \ref{p_bw_050_mc0.25} are
difficult to distinguish from the desired pressure and density
profiles throughout most of the star.  We also note that the
hydrodynamic and gravitational accelerations are very nearly equal in
magnitude and opposite in direction, as necessary for hydrostatic
equilibrium.

\subsection{Binary Equilibrium Configurations\label{binary}}
\subsubsection{SPH Calculations\label{binary_sph}}
The ability of our code to model close and contact binary systems for
thousands of orbits or longer is presented in
\cite{2009arXiv0904.0997G}.  Here we present our methods for modeling
equilibrium sequences of twin binaries, that is, binaries that consist
of two identical stars in synchronized orbit.  First, we place
identical relaxed stellar models along the $x$ axis with their centers
of mass separated by $r$.  
While the relaxation of the
binary takes place, the entropic variable particle values are held
constant.    The center of mass of the
entire system is fixed at the origin.
In addition,
the positions of the particles are continuously adjusted (by a simple uniform
translation along the binary axis), so that the separation between the
centers of mass equals the desired separation $r$.

The orbit is chosen to occur in the $xy$ plane.  The angular velocity
$\Omega_{\rm orb}$ defining the corotating frame is updated at every
timestep so that the centrifugal and inertial accelerations acting on
the fluid cancel.  In particular, we wish to find configurations in
which $\Omega_{\rm orb}^2(x_i {\rm \hat x} + y_i {\rm \hat y})=-(\dot
v_{x,i}{\rm \hat x}+\dot v_{y,i}{\rm \hat y}$), where the velocity
derivatives on the right hand side are components of acceleration in
the inertial frame.  By taking the dot product with $m_i(x_i{\rm \hat
x}+y_i{\rm \hat y})$ and then summing over all particles, we obtain
\begin{equation}
\Omega_{\rm orb}^2={-\sum_i m_i (x_i{\dot v}_{x,i}+y_i{\dot v}_{y,i})\over \sum_i m_i (x_i^2+y_i^2)}.
\end{equation}

We also include a drag force that opposes the velocity and provides a
contribution to the acceleration of $-{\bf v}_i/t_{\rm relax}$.  We
use $t_{\rm relax}=3$, approximately the fundamental period of
oscillation for our parent models.  We do not include any artificial
viscosity contribution when finding binary equilibrium configurations.

In order to find equilibrium configurations for a precisely equal mass
binary, even for configurations unstable to mass transfer, we enforce
a symmetry in particle properties: for each particle $i$ in star 1,
there is a partner particle $j$ in star 2 at $x_j=-x_i$ and $y_j=-y_i$
with velocity components $v_{x,j}=-v_{x,i}$, $v_{y,j}=-v_{y,i}$ and
with acceleration components ${\dot v}_{x,j}=-{\dot v}_{x,i}$, ${\dot
v}_{y,j}=-{\dot v}_{y,i}$.  All other properties are identical for any
such pair of particles.

The separation $r$ between the centers of mass can be
allowed to drift slowly so that an equilibrium
sequence is constructed: a so-called ``scanning run.''  In practice, we start runs that will scan
over separations by holding the centers of mass fixed at an initial
separation $r(0)$ for 40 time units, allowing the system to approach a
tidally bulged equilibrium configuration.  At an additional amount of
time $t$, the separation is set according to $r(t)=r(0) \left[
r(t_{scan})/r(0)\right]^{t/t_{scan}}.$ This form for $r(t)$ allows the
change in $r$ to occur at a decreasing rate as the stars approach and
interact more strongly, although the exact form is not critical to our
results.  We typically use $t_{scan}=300$, $r(0)=3.3$, and
$r(t_{scan})=2.1$.

\subsubsection{Data Reduction Methods\label{data}}

Once an SPH binary calculation has completed, we analyze the system at
various separations along the sequence.  To this end, a useful
quantity to consider is the effective potential, calculated as
\begin{equation}
\Phi_e(x,y,z)=\Phi(x,y,z)-\frac{1}{2}\Omega_{\rm orb}^2(x^2+y^2),\label{phie}
\end{equation}
where $\Phi$ is the gravitational potential, the coordinate $y$
measures perpendicular to the binary axis in the orbital plane, and
$z$ measures parallel to the rotation axis.  Along the binary axis
($y=z=0$), the effective potential has a local maximum $\Phi_e^{(i)}$
at $x=0$ (the inner Lagrangian point) and global maxima $\Phi_e^{(o)}$
at $|x|=x_o$ (the outer Lagrangian points). There are two minima at
$|x|=x_c$, corresponding to the cores of the two
components.\footnote{Note that in general $r\ne2x_c$, because we
define the binary separation $r$ as the distance between the {\em
centers of mass\/} of the two components.}  In equilibrium, the fluid
will fill up the effective potential well to some maximum, constant
level $\Phi_e^{(s)}$.  Borrowing the terminology from models of W UMa
binaries \citep[][and references therein]{ruc92}, we follow
\cite{1995ApJ...438..887R} and define the degree of contact $\eta$ as
\begin{equation}
\eta\equiv\frac{\Phi_e^{(s)}-\Phi_e^{(i)}}{\Phi_e^{(o)}-\Phi_e^{(i)}}. \label{eta}
\end{equation}
Clearly, we have $\eta<0$ for detached configurations: that is, none
of the fluid has a large enough effective potential energy to exceed
the effective potential energy barrier at the inner Lagrangian point.
For $0<\eta<1$, the effective potential of the fluid near $x=0$ does
exceed the barrier, and the system is classified as a contact binary.
For $\eta>1$, the envelopes overflow beyond the outer Lagrangian
surface, and no dynamical equilibrium configuration can be achieved;
that is, the system has exceeded the Roche limit.
Calculations in which we slowly scan to smaller separations can therefore determine position of first contact
($\eta=0$), the secular stability limit (at the minimum energy and
angular momentum along the sequence), and the Roche limit ($\eta=1$).

It is important to realize that the equilibrium sequence of a twin
binary passes smoothly from detached to contact configurations as the
separation $r$ decreases.  This is in contrast to all binary
equilibrium sequences with mass ratio $q\ne 1$, which terminate at a
Roche limit corresponding to the onset of mass transfer through the
inner Lagrangian point (that is, once one of the binary components
overflows its Roche lobe).  For twin binaries, however, the Roche
limit, which we still define as the last equilibrium configuration
along a sequence with decreasing $r$, corresponds to the onset of mass
shedding through the outer Lagrangian points: as an example, note that
several particles have been shed to the far left and far right of the
$r=2.22$ frame in Fig.\ \ref{p6bw_040_170_245_260_297_310}.

We estimate $\Phi_e^{(s)}$ from our SPH models by finding the maximum
effective potential of the points along the $x$-axis that are within
one smoothing length of the center of an SPH particle.  Thus, even if
the centers of all SPH particles are within the outer Lagrangian
surface, we may still consider the system as having reached the Roche
limit when some smoothing kernels extend substantially beyond the
outer Lagrangian surface.  Such an estimate accounts for the fact that
an SPH particle is not a point mass but instead represents a parcel of
fluid with a density profile described by the smoothing kernel.  Our
means of estimating $\Phi_e^{(s)}$ allow our critical separation
results to converge quickly to a steady value as the
resolution is increased up to the resolution presented in this paper
(see \S\ref{equilibrium_model_sequences}).

\subsection{Dynamical Calculations\label{dynamical}}

We generate initial conditions for our dynamical runs by taking a
configuration at the desired separation $r$ from a scanning run and
then relaxing for an additional 200 time units.  If a particle escapes
past an outer Lagrangian point during this time interval, then we end
the relaxation stage and begin following the dynamics immediately. During dynamical calculations, we include no
drag force and move the particles according to their velocities in the
usual way \citep[for details, see][]{2009arXiv0904.0997G}.  Particles
are again treated independently so that mass transfer events can be
followed; that is, unlike the scans described in \S\ref{binary}, no
symmetry constraints are applied to particle motion.  Artificial
viscosity is implemented in both the acceleration and the entropy
equations.  To minimize the spurious effects of artificial viscosity
\citep{lsrs99}, our dynamical calculations are done in a rotating
frame, with the the angular velocity $\Omega_{\rm orb}$ calculated
once at the beginning of the dynamical evolution and thereafter held
constant when applying Coriolis and centrifugal forces.

%
%

\section{Results} \label{numerical_results}

Using the methods described above, we construct twin binary
sequences for 16 different core masses $m_c$ listed in Table
\ref{parents} and covering the range from 0 to 0.99.  In
\S\ref{equilibrium_model_sequences}, we 
create an equilibrium sequence
for each core mass by slowly scanning over the binary separation, thereby identifying the separations of first contact, of
the secular instability (if it exists), and of the Roche limit.  In
\S\ref{dynamical_integrations}, separate dynamical calculations of
various initial separation $r$ then allow us to test the dynamical
stability of the contact configurations and to follow any mass
transfer and the merger in unstable systems.


\subsection{Equilibrium Sequences\label{equilibrium_model_sequences}}

\begin{figure}
\includegraphics[width=3.4in]{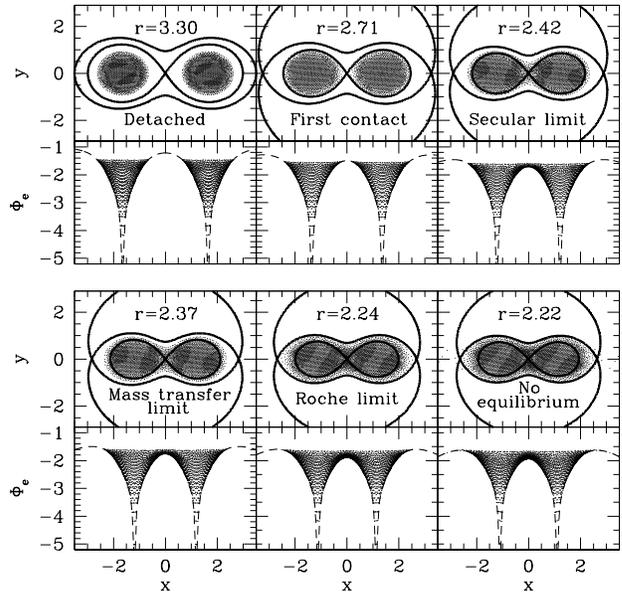}
\caption{Sequence of binary equilibrium configurations for two
identical condensed polytropes of core mass $m_c=0.1$. Projections
onto the orbital plane (the $xy$ plane) are shown at six different
binary separations for those SPH particles with $|z|<0.06$.  The thick
solid curves represent two surfaces of constant effective potential
$\Phi_e$ (see eq.~[\ref{phie}]): namely, the inner and outer
Lagrangian surfaces passing through the points L1 and L2.  Shown
beneath each configuration are corresponding projections onto the
$(x,\Phi_e)$ plane for the same particles.  The dashed curves give the
variation of $\Phi_e$ along the binary axis ($y=z=0$).  Contact
configurations are obtained when the binary separation $r\lesssim 2.71$
(in units where an isolated binary component has radius $R=1$).  For
$r\lesssim 2.24$, mass shedding through the outer Lagrangian points
occurs, and no equilibrium configuration exists.
\label{p6bw_040_170_245_260_297_310}
}
\end{figure}

Representative snapshots along the $m_c=0.1$ equilibrium sequence are
presented in Figure~\ref{p6bw_040_170_245_260_297_310}.  The structure
of these solutions is shown both in projection onto the orbital plane
(the $xy$ plane) and in terms of the effective potential $\Phi_e$.
The thick solid curves in Figure~\ref{p6bw_040_170_245_260_297_310}
are the surfaces of constant effective potential $\Phi_e$ that mark
the inner and outer Lagrangian surfaces.  For fixed $x$, $\Phi_e$ is
minimum on the binary axis ($y=z=0$), and this minimum value is given
as a dashed curve in Figure~\ref{p6bw_040_170_245_260_297_310}. In
hydrostatic equilibrium, the fluid fills up to a constant level
$\Phi_e^{(s)}$ that is independent of $x$.
%

\begin{figure}
\includegraphics[width=3.4in]{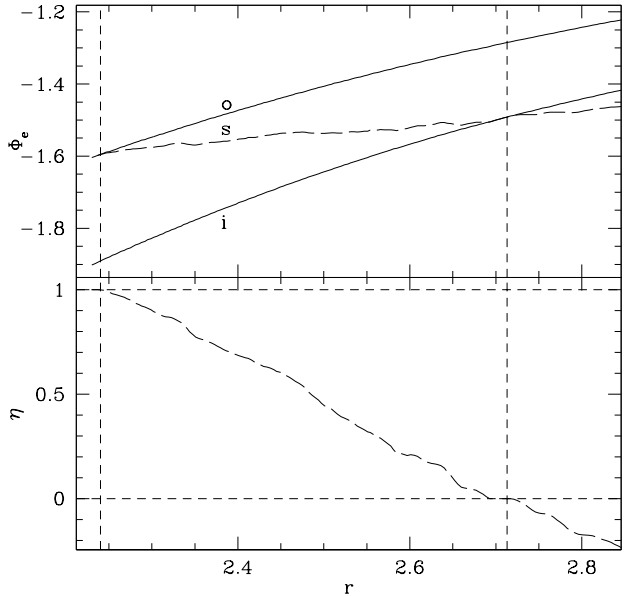}
\caption{Variation of critical effective potentials along the
  equilibrium sequence presented in
  Fig.~\ref{p6bw_040_170_245_260_297_310}.  Values of the effective
  potential at the outer Lagrangian surface (solid curve o), the inner
  Lagrangian surface (solid curve i), and the fluid surface (long
  dashed curve s) are shown as a function of binary separation $r$ in
  the top frame.  The degree of contact $\eta$ (eq.\ [\ref{eta}]) is
  shown in the bottom frame as the long dashed curve.  The short
  dashed curves give the positions of first contact ($\eta=0$) and of
  the Roche limit ($\eta=1$).
\label{fig:eta}
}
\end{figure}

Referring to Figure~\ref{p6bw_040_170_245_260_297_310} we see that at
the initial separation in our scan, $r=3.30$, the system is tidally
bulged and the binary is detached: the fluid does not extend out to
the inner Lagrangian surface (also known as the Roche lobe).  At a
separation $r=2.71$, the binary stars fill the inner Lagrangian
surface and make first contact through the L1 point, located at the
origin.  Once the separation decreases to $r=2.42$, the binary reaches
the secular instability limit, as marked by a minimum in the energy
$E$ and angular momentum $J$ along this equilibrium sequence (see
below). The critical separation $r=2.37$ at which mass transfer
commences is identified with dynamical calculations described in
\S\ref{dynamical_integrations}; in this and other scans through
equilibrium configurations, however, we suppress the mass transfer by
enforcing symmetry in particle properties (see \S\ref{binary_sph}).
At a separation $r=2.24$, the fluid extends out to the outer
Lagrangian surface, marking the Roche limit.  At even smaller
separations, for example at the separation $r=2.22$ shown in the
figure, no equilibrium configurations exist and the stars shed mass
through the outer Lagrangian surface near the L2 point.  The variation
of critical effective potentials and the degree of contact $\eta$
along this $m_c=0.1$ equilibrium sequence is illustrated in
Figure~\ref{fig:eta}.  For future reference, we note the approximately
linear dependence of $\eta$ on the separation $r$.

\begin{figure}
\includegraphics[width=3.4in]{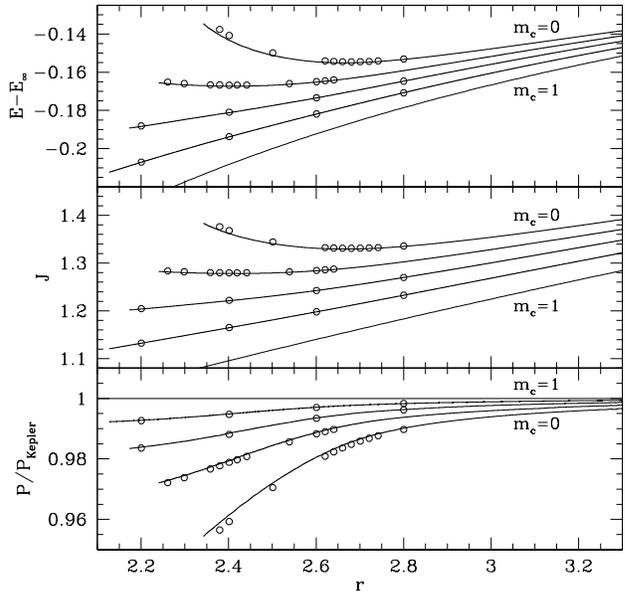}
\caption{Variation of the system energy $E$ (relative to the total
self-energy $E_\infty$ of the binary components at infinity), angular
momentum $J$, and orbital period $P$ along the equilibrium sequence
for twin binaries with $m_c=0$, 0.1, 0.25, 0.5, and 1.  In the $E$ and
$J$ frames, higher curves correspond to smaller core mass, while in
the $P$ frame higher curves correspond to larger core mass.  The
orbital period $P$ is normalized to the analytic point mass result
$P_{\rm Kepler}=2^{1/2}\pi r^{3/2}$.  The curves are from SPH scans of the
equilibrium sequence, except for the $m_c=1$ curve which is the
analytic result for two point masses.  The $m_c=0$ and $0.1$ curves
exhibit a minimum in $E$ and $J$, marking the position of the secular
instability limit.  The curves from the SPH scans terminate at the
Roche limit, where mass shedding through the outer Lagrangian point
commences.  Critical points along our SPH scans are presented in Tables \ref{firstcontact}, \ref{secular}, and \ref{roche}.
The individual data points (open circles) in this figure result from relaxing for an
additional 200 time units at the given separation $r$.  The agreement
between these points and their corresponding scan helps to confirm
that our scans are indeed producing equilibrium sequences.
\label{vsrallwithp}
}
\end{figure}

From Figure~\ref{vsrallwithp}, we note that as $m_c$ increases toward
1, our results approach the Keplerian solution of two orbiting point
masses.  As expected, deviations from the point mass result increase
as a given binary becomes more deeply in contact or as we consider a
binary associated with a smaller core mass.  From the bottom frame of
Figure~\ref{vsrallwithp}, we note that smaller core masses
(corresponding to stars with less centrally concentrated density
profiles) have a smaller orbital period at any given separation.  For
$m_c<1$, tidal interactions between the two stars make the effective
potential stronger than $1/r$ and shorten the rotation period compared
to a point mass system.  For $m_c=0.1$, for example, the deviation of
the orbital period from the point mass result is approximately 1\% at
first contact, 2\% at the secular instability limit, and 3\% at the
Roche limit.

As in \cite{1995ApJ...438..887R}, we determine the secular stability
limit along the equilibrium sequence by locating where both the total
energy $E$ and total angular momentum $J$ are minimum in curves such
as those of Figure~\ref{vsrallwithp}.  Our numerical results provide an
accurate determination of this point for a close binary system, as the
separate minima in $E$ and $J$ coincide to high numerical
accuracy. This is in accord with the general property that
$dE=\Omega_{\rm orb}\,dJ$ along any sequence of uniformly rotating fluid
equilibria \citep{1969ApJ...157.1395O}.

\begin{figure}
\includegraphics[width=3.4in]{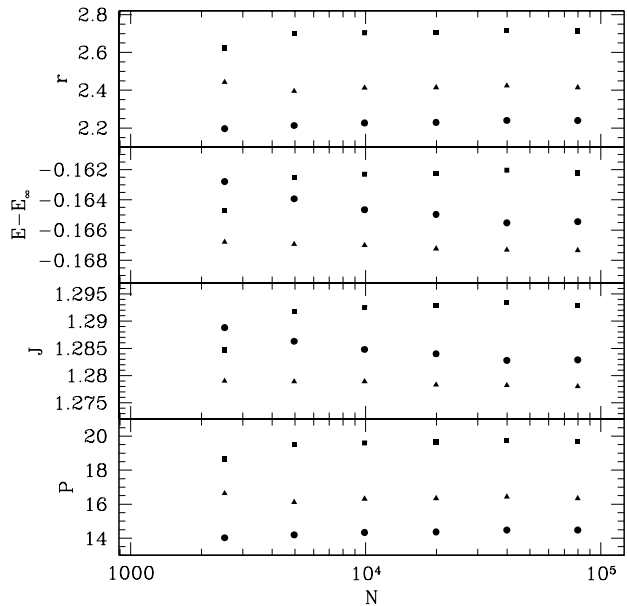}
\caption{Critical separation $r$, energy $E-E_\infty$, angular
momentum $J$, and orbital period $P$ at first contact (squares),
secular instability (triangles), and the Roche limit (circles) versus
total particle number $N$ for several $m_c=0.1$ equilibrium scans.
Note the convergence of results for $N\gtrsim 10^4$.  Most of the
binary calculations in this paper employ $N\approx 4\times 10^4$
particles.
\label{varyN}
}
\end{figure}

For the $m_c=0$ binary, secular instability occurs soon after contact
along this sequence and therefore stable, long-lived equilibrium
configurations can exist only in shallow contact, $\eta\lesssim 0.2$.
In contrast, the sequences with non-zero core masses permit much
deeper contact before the secular instability is reached.  For
example, a binary with core masses of $0.125$ does not reach the
secular instability until nearly $\eta=0.9.$ For core masses
$m_c\gtrsim 0.15$, the stars will reach the Roche limit before the
secular instability limit.  Tables~\ref{firstcontact}, \ref{secular},
and \ref{roche} respectively present system properties at first
contact ($\eta=0$), the secular instability limit (where $E$ and $J$
are minima), and at the Roche limit ($\eta=1$) for our sequences of
various core mass $m_c$.  Additional runs at varying resolution
indicate that the results in our tables have converged to within $\sim
1$\% (e.g.\ see Fig.\ \ref{varyN}).

\begin{table}[ht!]
\caption{FIRST CONTACT ALONG THE EQUILIBRIUM SEQUENCES
          OF TWIN BINARIES\tablenotemark{a}}
\begin{tabular}{ccccc} \tableline\tableline
   $m_c$  &  $r$   &     $P$   &$E-E_\infty$& $J$\\
\tableline
   0.000&  2.75  &        20.1       & -0.154   & 1.33\\
   0.050&  2.73  &        19.8       & -0.159   & 1.31\\
   0.100&  2.72  &        19.7       & -0.162   & 1.29\\
   0.125&  2.72  &        19.8       & -0.163   & 1.29\\
   0.150&  2.71  &        19.7       & -0.165   & 1.28\\
   0.175&  2.69  &        19.5       & -0.167   & 1.27\\
   0.200&  2.69  &        19.5       & -0.168   & 1.27\\
   0.250&  2.69  &        19.5       & -0.170   & 1.25\\
   0.300&  2.70  &        19.6       & -0.171   & 1.25\\
   0.400&  2.68  &        19.4       & -0.175   & 1.23\\
   0.500&  2.66  &        19.3       & -0.178   & 1.21\\
   0.600&  2.67  &        19.3       & -0.180   & 1.20\\
   0.700&  2.66  &        19.2       & -0.183   & 1.18\\
   0.800&  2.67  &        19.4       & -0.184   & 1.17\\
   0.900&  2.67  &        19.4       & -0.186   & 1.16\\
   0.990&  2.70  &        19.7       & -0.185   & 1.16\\
\tableline
\end{tabular}
\tablenotetext{a}{Units are defined such that $G=M=R=1$, $m_c$ is the core mass of each component, $r$ is the binary separation,
$\eta$ is the degree of contact (eq.\ [\ref{eta}]), $P$ is the orbital period, and $E-E\infty$ and $J$ are the
orbital energy and angular momentum,
respectively; the energy $E_\infty$ is the total equilibrium energy at
infinite separation (that is, twice the energy of a single component
in isolation).}
\label{firstcontact}
\end{table}

\begin{table}
\caption{SECULAR INSTABILITY ALONG THE EQUILIBRIUM SEQUENCES
          OF TWIN BINARIES\tablenotemark{b}}
\begin{tabular}{cccccc}  \tableline\tableline
   $m_c$  &  $r$   &  $\eta$  &   $P$   &$E-E_\infty$& $J$\\
\tableline 
   0.000&  2.67  &   0.17   &    19.1       & -0.155   & 1.33\\
   0.050&  2.55  &   0.40   &    17.7       & -0.161   & 1.30\\
   0.100&  2.42  &   0.64   &    16.4       & -0.167   & 1.28\\
   0.125&  2.29  &   0.86   &    15.1       & -0.171   & 1.26\\
   0.150&  2.22  &   1.0    &    14.3       & -0.175   & 1.25\\
\tableline
\end{tabular}
\tablenotetext{b}{In twin binary sequences with core masses $m_c\gtrsim 0.15,$ the Roche limit is reached before
the secular limit.  Units and column headings are as in Table 2, footnote a; the degree of contact $\eta$ is defined by eq.~\ref{eta}.}
\label{secular}
\end{table}

\begin{table}
\caption{ROCHE LIMIT\tablenotemark{c} ALONG THE EQUILIBRIUM SEQUENCES
          OF TWIN BINARIES\tablenotemark{d} }
\begin{tabular}{cccccc} \tableline\tableline
   $m_c$  &  $r$    &   $P$   &$E-E_\infty$& $J$\\
\tableline
   0.000&  2.34  &   15.2       & -0.135   & 1.38\\
   0.050&  2.28  &   14.7       & -0.153   & 1.32\\
   0.100&  2.24  &   14.5       & -0.166   & 1.28\\
   0.125&  2.22  &   14.3       & -0.171   & 1.27\\
   0.150&  2.21  &   14.3       & -0.175   & 1.25\\
   0.175&  2.20  &   14.2       & -0.179   & 1.24\\
   0.200&  2.19  &   14.1       & -0.183   & 1.22\\
   0.250&  2.17  &   14.0       & -0.189   & 1.20\\
   0.300&  2.16  &   13.9       & -0.195   & 1.18\\
   0.400&  2.14  &   13.8       & -0.204   & 1.15\\
   0.500&  2.13  &   13.7       & -0.212   & 1.12\\ 
   0.600&  2.12  &   13.6       & -0.219   & 1.10\\
   0.700&  2.12  &   13.6       & -0.224   & 1.08\\
   0.800&  2.10  &   13.5       & -0.231   & 1.05\\
   0.900&  2.10  &   13.5       & -0.235   & 1.04\\
   0.990&  2.12  &   13.7       & -0.236   & 1.03\\
\tableline
\end{tabular}
\tablenotetext{c}{Defined as the equilibrium configuration with the
minimum binary separation.}
\tablenotetext{d}{Unit and column definitions are identical to those in Table 2, footnote a.}
\label{roche}
\end{table}

\begin{figure}
\includegraphics[width=3.4in]{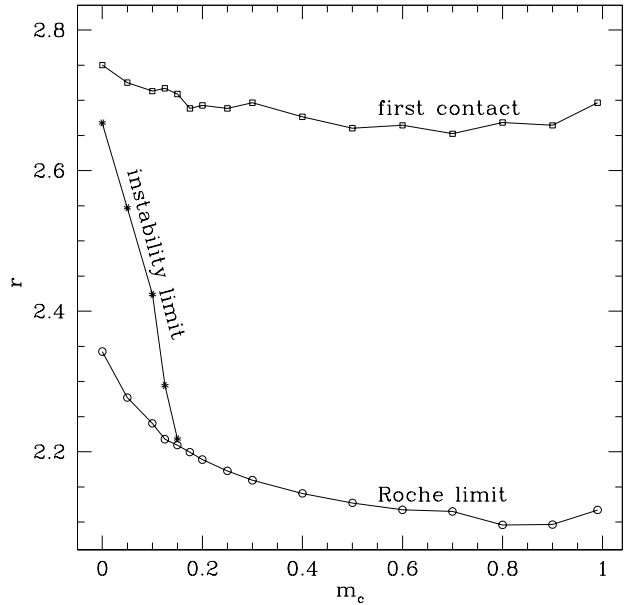}
\caption{Separation at first contact (squares), the onset of secular
instability (stars), and the Roche limit (circles) versus core mass
$m_c$.  The data points are determined from SPH scans of equilibrium sequences, as reported in Tables \ref{firstcontact}, \ref{secular}, and \ref{roche}.  The lines are simply to help guide the eye.
\label{rcrit}
}
\end{figure}

These results for critical separations are summarized in
Figure~\ref{rcrit}.  Due to tidal effects, the volume of each star is
typically 1-2\% larger at first contact than it is for that star in
isolation.  The separation $r_{\rm fc}$ at first contact is only
weakly dependent on $m_c$, being within 2\% of 2.7 for any core mass.
For comparison, we note that the standard simple treatment of twin
binaries would imply a first contact separation of $1/0.3799=2.63$,
where the factor 0.3799 comes from numerical integration of Roche lobe
volumes around identical point masses
\citep[e.g.,][]{1983ApJ...268..368E} and any change in the volumes of
the stars due to tidal effects is neglected.  We see therefore that
finite size effects act to increase the separation of first contact.

All three of the critical separations considered (first contact,
secular instability, and Roche limit) tend to decrease as the core
mass increases.  It is straightforward to find fitted formulas for the
critical separations that are accurate to within $\sim 1$\% for any
core mass $m_c$: for first contact
\begin{equation}
r_{\rm fc}\approx 2.66+0.08(1-m_c)^4\label{rfc}
\end{equation}
and for the secular instability limit
\begin{equation}
r_{\rm sec}\approx 2.69-3m_c.\label{rsec}
\end{equation}
We note that the $1-m_c$ in equation (\ref{rfc}) equals the envelope mass (in units where the total stellar mass $M=1$).
We give our fit for the Roche limit
separations in the next subsection, where we can determine these data with
slightly better accuracy.  We have not fit for the slight increase in the
first contact data as the core mass is increased from
$m_c= 0.9$ to 0.99, as this feature is a numerical artifact due to our
$m_c=0.99$ single star equilibrium model settling to a radius a few
percent larger than 1.

The critical core mass $m_c\approx 0.15$, for which the secular
instability and Roche limits coincide, can be determined
graphically from Figure~\ref{rcrit} by extrapolating the line
connecting the secular instability data down to the Roche limit curve.
This intersection is important because it implies that all of our
equal mass binaries with $m_c\gtrsim 0.15$ can stably exist in deep
contact, at separations all the way down to the Roche limit.  Thus,
essentially all twin red giant binaries will coalesce only due to mass
shedding through the outer Lagrangian points.  In contrast, twin
binaries with $m_c\lesssim 0.15$, corresponding primarily to
main-sequence stars and subgiants, reach the secular instability limit
at a larger separation than that of the Roche limit.  As we will see
in the next subsection, the secular instability limit is usually
accompanied by {\it dynamical} mass transfer at the same or a slightly
smaller separation.  Therefore, main sequence and subgiant twin
binaries, as contrasted to most red giant twins, will start coalescing (a) when in
more shallow contact and (b) through mass transfer across the inner
Lagrangian point.

\subsection{Dynamical Integrations\label{dynamical_integrations}}

\begin{figure}
\includegraphics[width=3.4in]{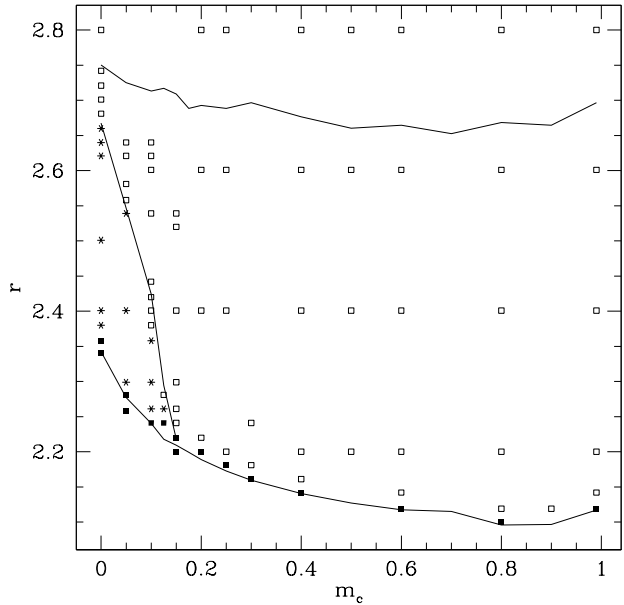}
\caption{Results of dynamical integrations: stable binaries (open
  squares), dynamically unstable binaries (asterisks), and configurations
  with no equilibrium (filled squares).  The curves represent the
  critical separations as determined by scanning runs, as in
  Fig.~\ref{rcrit}.  The agreement between the results of the
  equilibrium sequences and dynamical calculations is excellent.
\label{aa_20000}
}
\end{figure}

We now study the stability of binary configurations with fully
dynamical SPH integrations (see \S\ref{dynamical} for details of the
setup).  Figure~\ref{aa_20000} summarizes the results of nearly 100
dynamical simulations with various $m_c$ and initial $r$ values.\footnote{Visualizations of selected simulations are available at http://webpub.allegheny.edu/employee/j/jalombar/movies/.}
We find the Roche limit to be very nearly at the separations determined
from the equilibrium scans, and the additional relaxation that we
perform before beginning a dynamical calculation allows us to
determine these separations even more accurately.  In addition, we
find that most systems that are secularly unstable are also
dynamically unstable to mass transfer and then merger, as discussed below.

\begin{figure}
\includegraphics[width=3.4in]{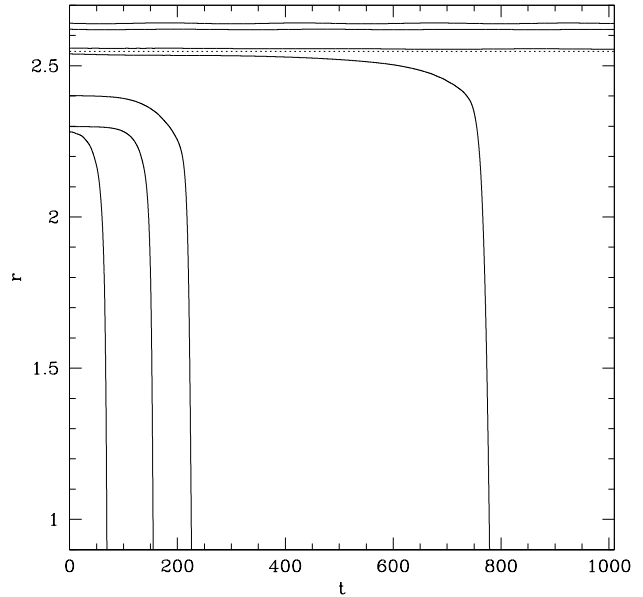}
\includegraphics[width=3.4in]{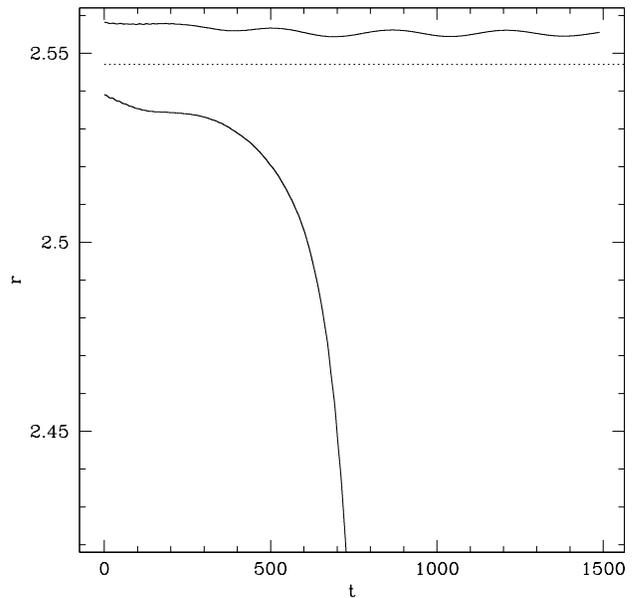}
\caption{{\it Top plot}: Separation $r$ versus time in several different
dynamical SPH calculations for the core mass $m_c=0.05$ twin binaries.
The initial separations are $r = 2.64$, 2.62,
2.56, 2.54, 2.40, 2.30 and 2.28. {\it Bottom plot:} Zoomed in view of
the $r=2.56$ and 2.54 cases, which straddle the instability limit.
In both plots, the horizontal dotted line represents the
secular instability limit at $r=2.547$,
as identified by an equilibrium scan.
\label{rvst0.05}
}
\end{figure}

The time evolution of the separation of $m_c=0.05$ 
twin binaries is illustrated in
Figure~\ref{rvst0.05} 
for several different initial separations.  This figure
also indicates the secular
instability limit $r_{\rm sec}=2.547$, as determined by the energy and
angular momentum minima in the equilibrium sequence of these binaries
(see \S\ref{equilibrium_model_sequences}).
Systems with separations
$r\gtrsim r_{\rm sec}$ are clearly {\it dynamically} stable, while
those with $r\lesssim r_{\rm sec}$ are {\it dynamically}
unstable.  That is, the secular and
dynamical stability limits coincide or very nearly coincide,
at least at this core mass.

The bottom plot of Figure \ref{rvst0.05} shows the dynamical evolution
of two cases that straddle the instability limit.  For the $r=2.56$
case, an
epicyclic period of 350 time units is clearly evident,
much larger than the orbital period of 17.9 time units. The
large difference in these periods is an
indication of how close the system is to an instability limit
\citep{1994ApJ...432..242R}.  Indeed, if $r$ were precisely at the
dynamical stability limit, the period of small epicyclic oscillations
would formally be infinite.

\begin{figure*}
~~~\includegraphics[angle=270,width=5in]{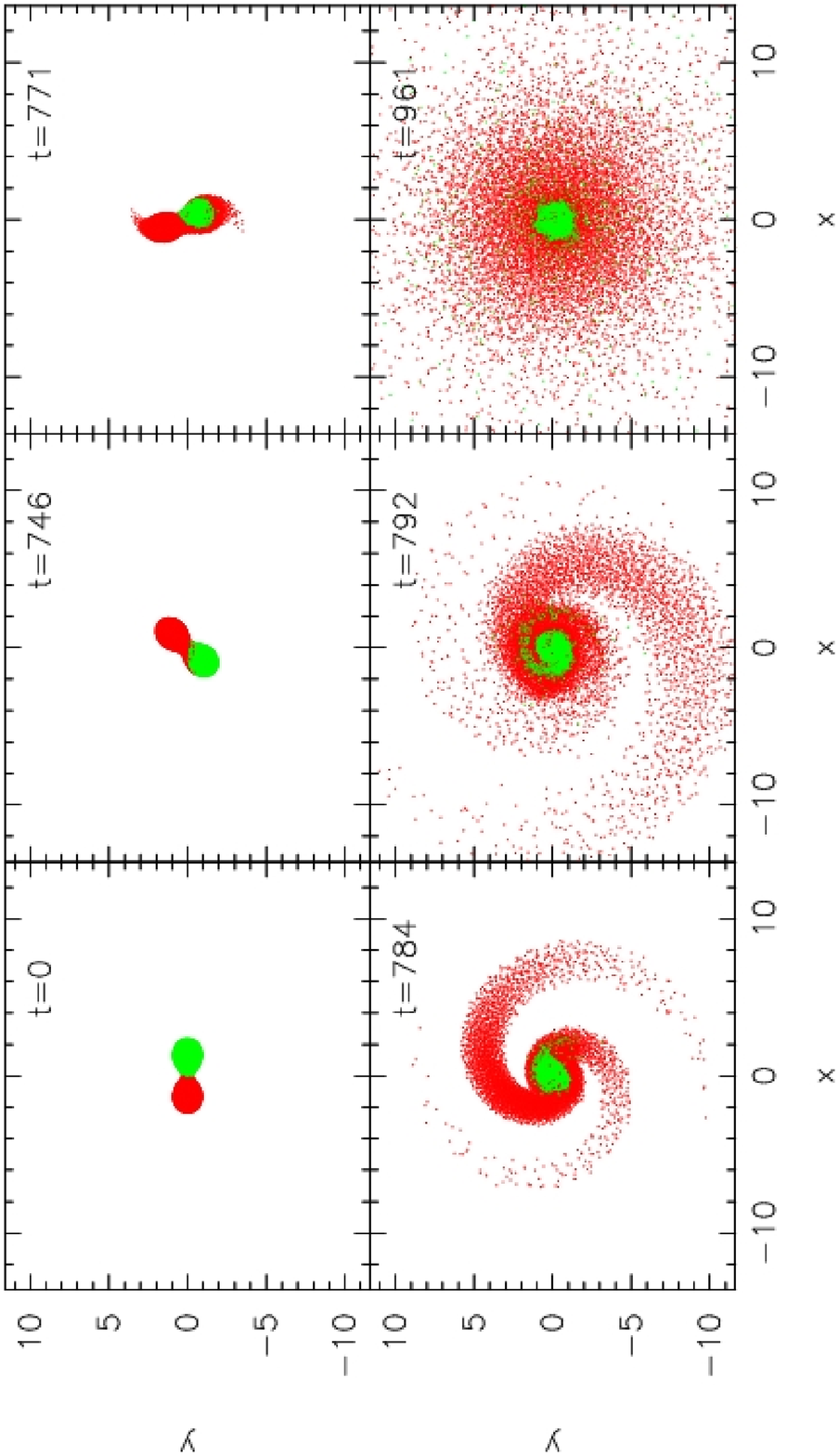}\\~\\
\includegraphics[angle=270,width=6in]{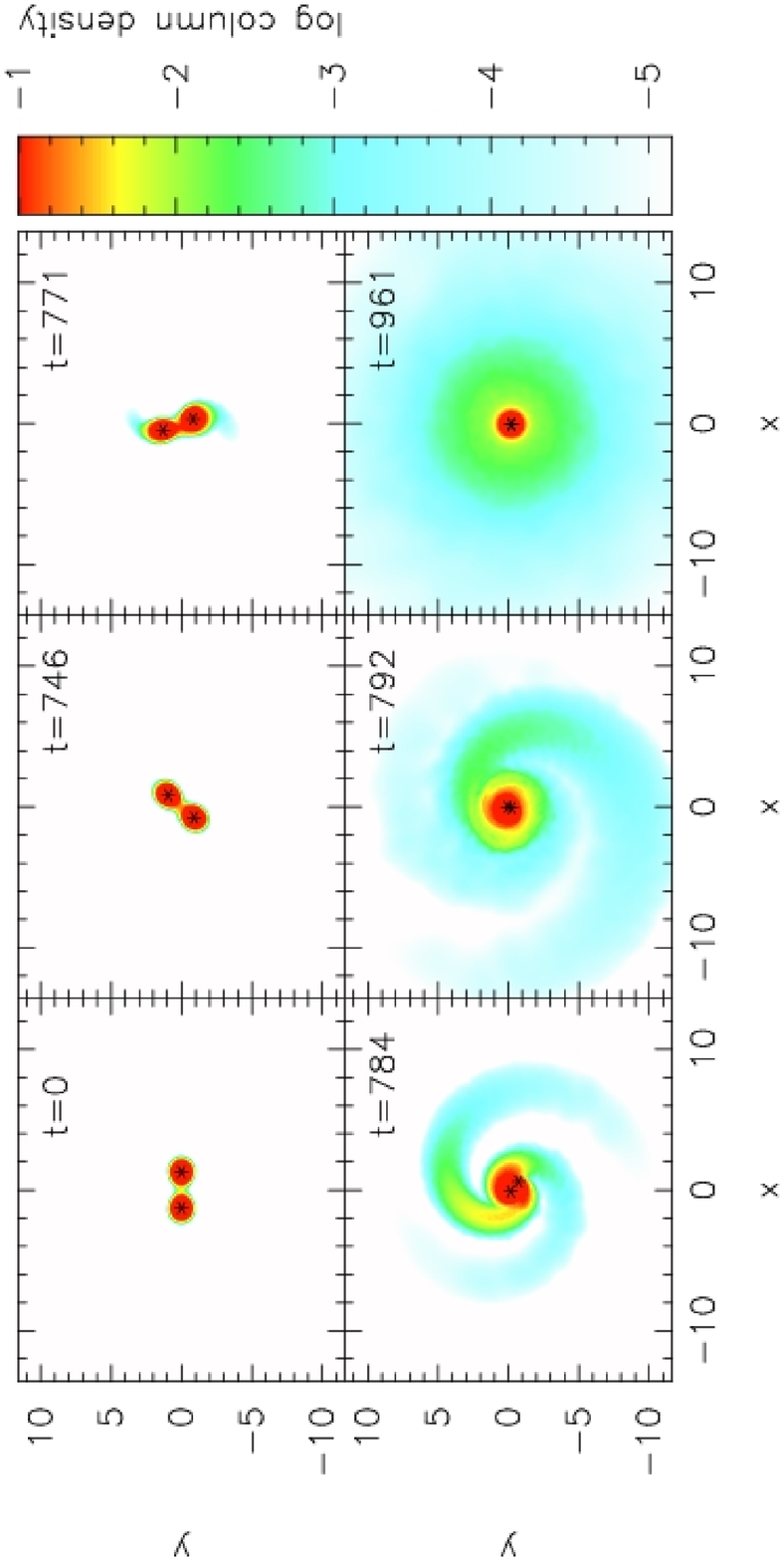}
\caption{{\it Top frames:} Projection of SPH particles onto the orbital
  plane at six different times in the merger of a $m_c=0.05$ binary
  with initial separation $r=2.54$.  Particles are colored according
  to the star in which they originated.  {\it Bottom frames:} Column
  density plots at the same six moments, with the asterisks
  representing the positions of the compact cores.
\label{fs0.050_r2.54b}
}
\end{figure*}

Figure~\ref{fs0.050_r2.54b} presents projected particle positions (top)
and column density plots (bottom) at six different times in the
$m_c=0.05$, $r=2.54$ dynamical calculation, a case just inside the
instability limit. Colors in the particle plot are used to indicate
from which component the particles originated.  The coordinate system
used here rotates counterclockwise with a period of 17.64 time units,
which equals the orbital period of this binary at early times, before
significant mass transfer has occurred.  The instability initially
manifests itself in the form of a narrow arm of gas that begins in the
outer layers of one star, gradually flows across the neck surrounding
the inner Lagrangian point, and then creeps around the other star (see
the $t=746$ and 771 particle plot frames).  The mass transfer drives
the binary components closer, triggering the excretion of mass through
the outer Lagrangian points ($t=771$) and accelerating the inspiral of
the two cores.  By $t=784$, the mass transfered from one star has
completely engulfed the other.  At later times, the merger product
approaches a rapidly rotating, axisymmetric configuration centered on
the two cores orbiting in a tight binary (see the $t=792$ and 961
frames).

\begin{figure}
\includegraphics[width=3.4in]{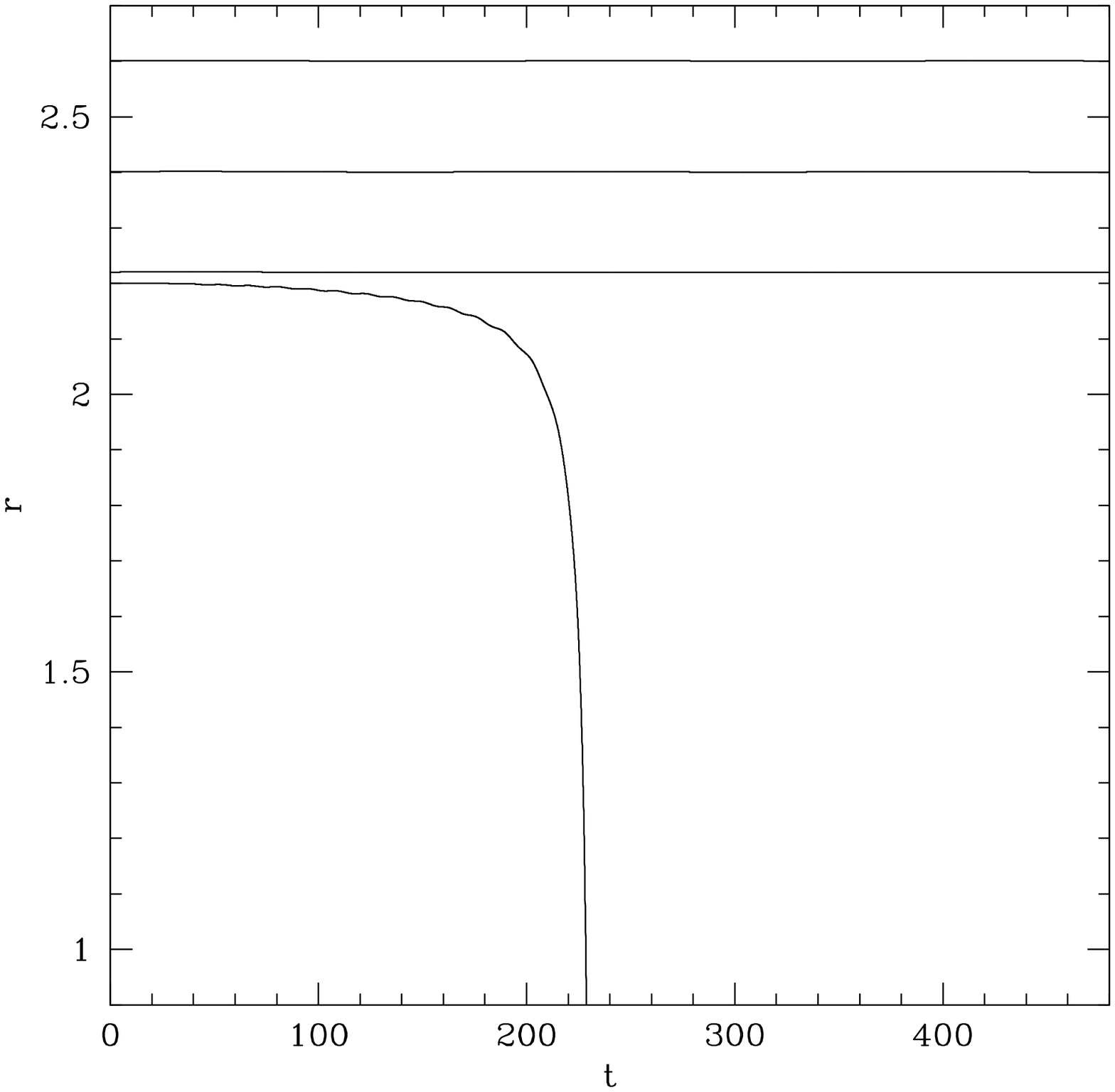}
\includegraphics[width=3.4in]{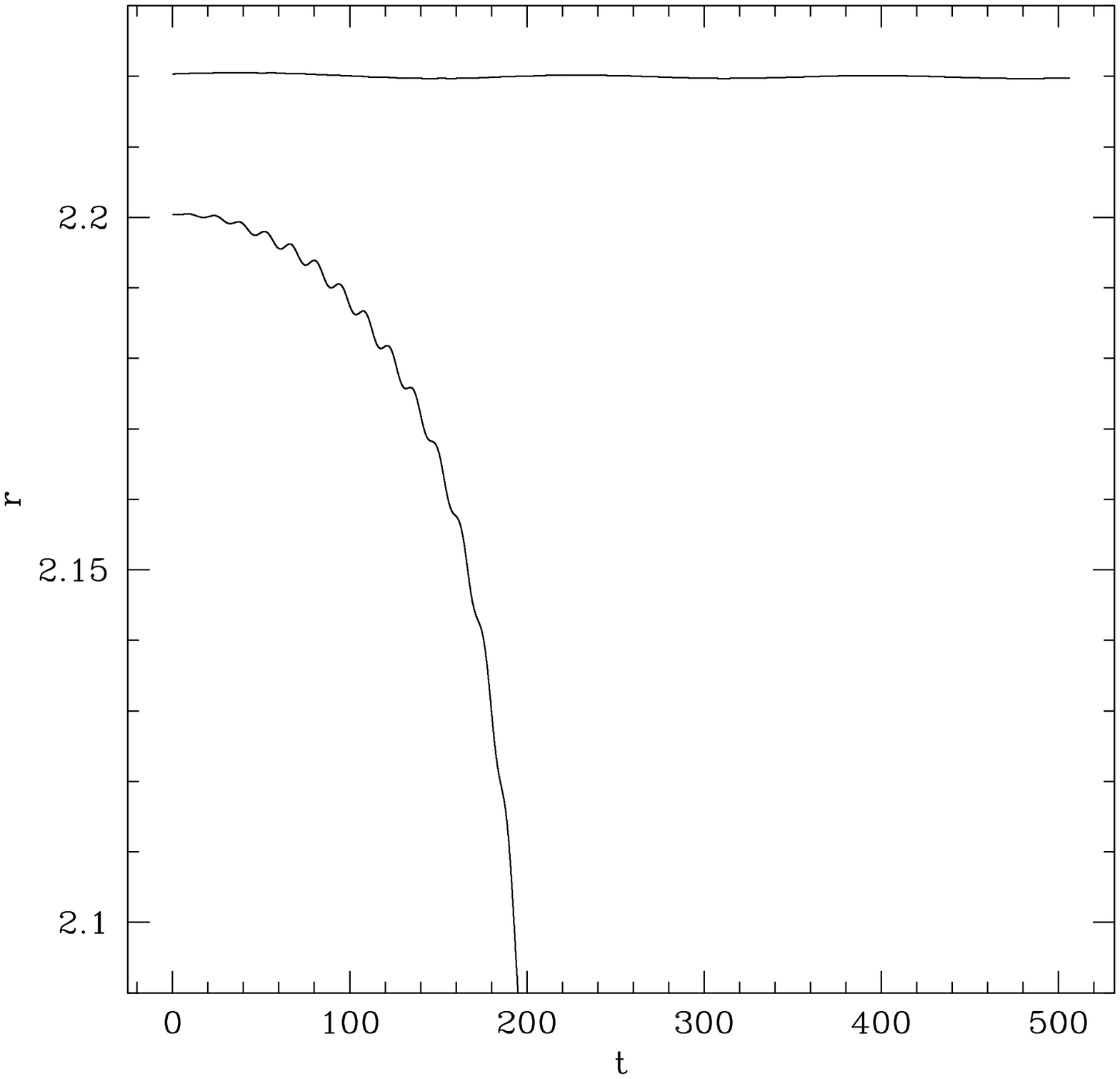}
\caption{Like Fig.~\ref{rvst0.05}, but for twin binaries with core
  mass $m_c=0.2$.  At this core mass, there is no secular instability
  limit: instead, the system become unstable only once it reaches
  the Roche limit at $r\approx 2.2$.
The initial separations shown in the top plot are $r=2.60$,
2.40, 2.22, 2.20, while a zoomed in view of the $r=2.22$ and 2.20
curves are shown in the bottom plot.
\label{rvst0.2}
}
\end{figure}

Figure~\ref{rvst0.2} shows the evolution of binary separation for
$m_c=0.2$ cases.  Recall from \S\ref{equilibrium_model_sequences} that
there is no secular instability limit at this core mass.  Instead,
stable equilibrium models exist all the way to the Roche limit.  A
binary at an initial separation $r=2.22$ (or larger) orbits stably.
In contrast, a binary at $r=2.20$ gradually loses mass through the
outer Lagrangian points, triggering a stage of rapid coalescence.  The
period of mass loss persists for several orbital periods: each of the
oscillations superposed on the decreasing $r=2.20$ curve in the bottom
plot of Fig.\ref{rvst0.2} corresponds to one orbital period.  These
dynamical calculations indicate that the Roche limit for a $m_c=0.2$
twin binary indeed occurs near $r=2.2$, in excellent agreement with
the $r=2.19$ critical value estimated from the equilibrium scan.

\begin{figure*}

~~~\includegraphics[angle=270,width=5in]{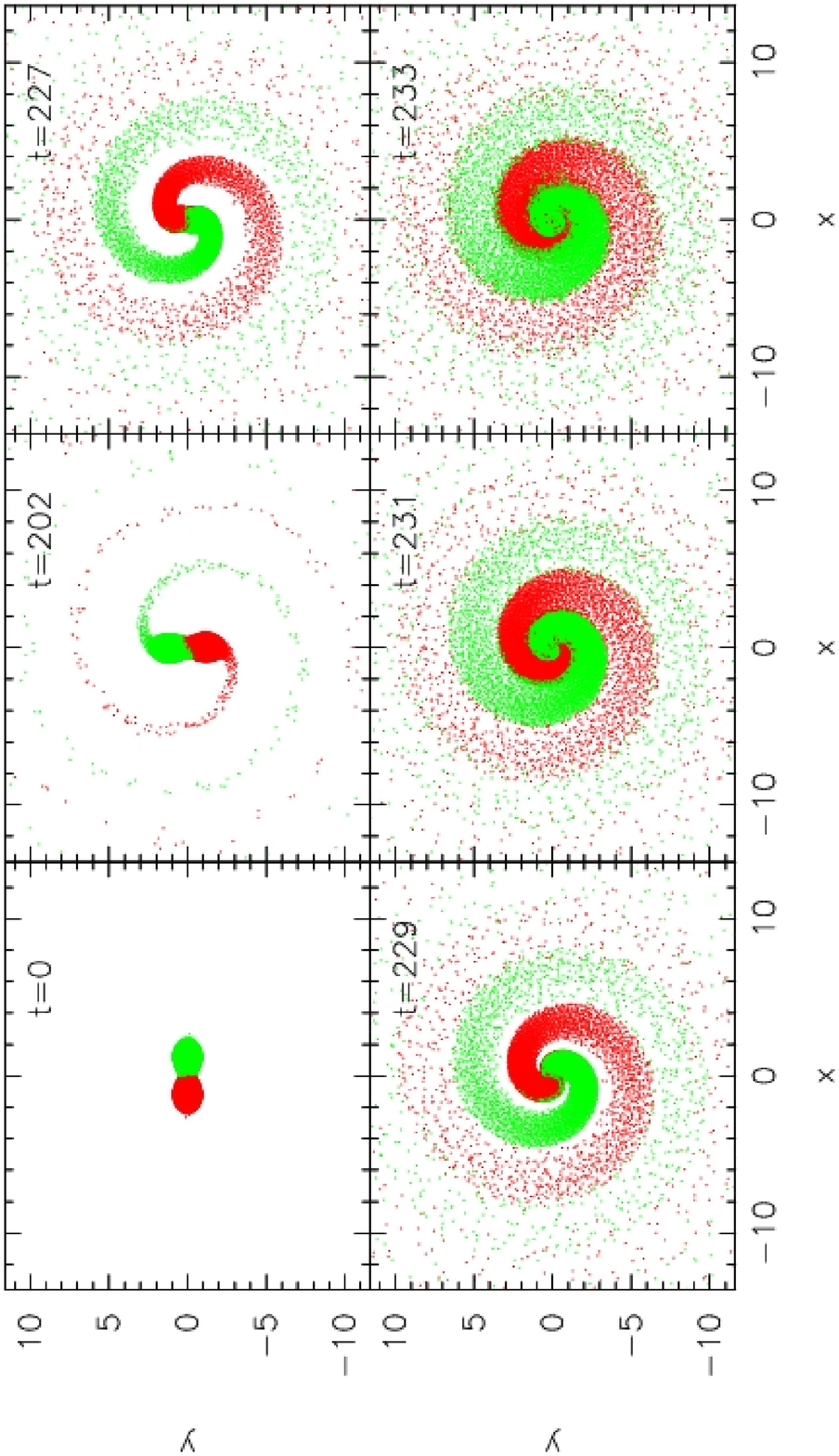}\\~\\
\includegraphics[angle=270,width=6in]{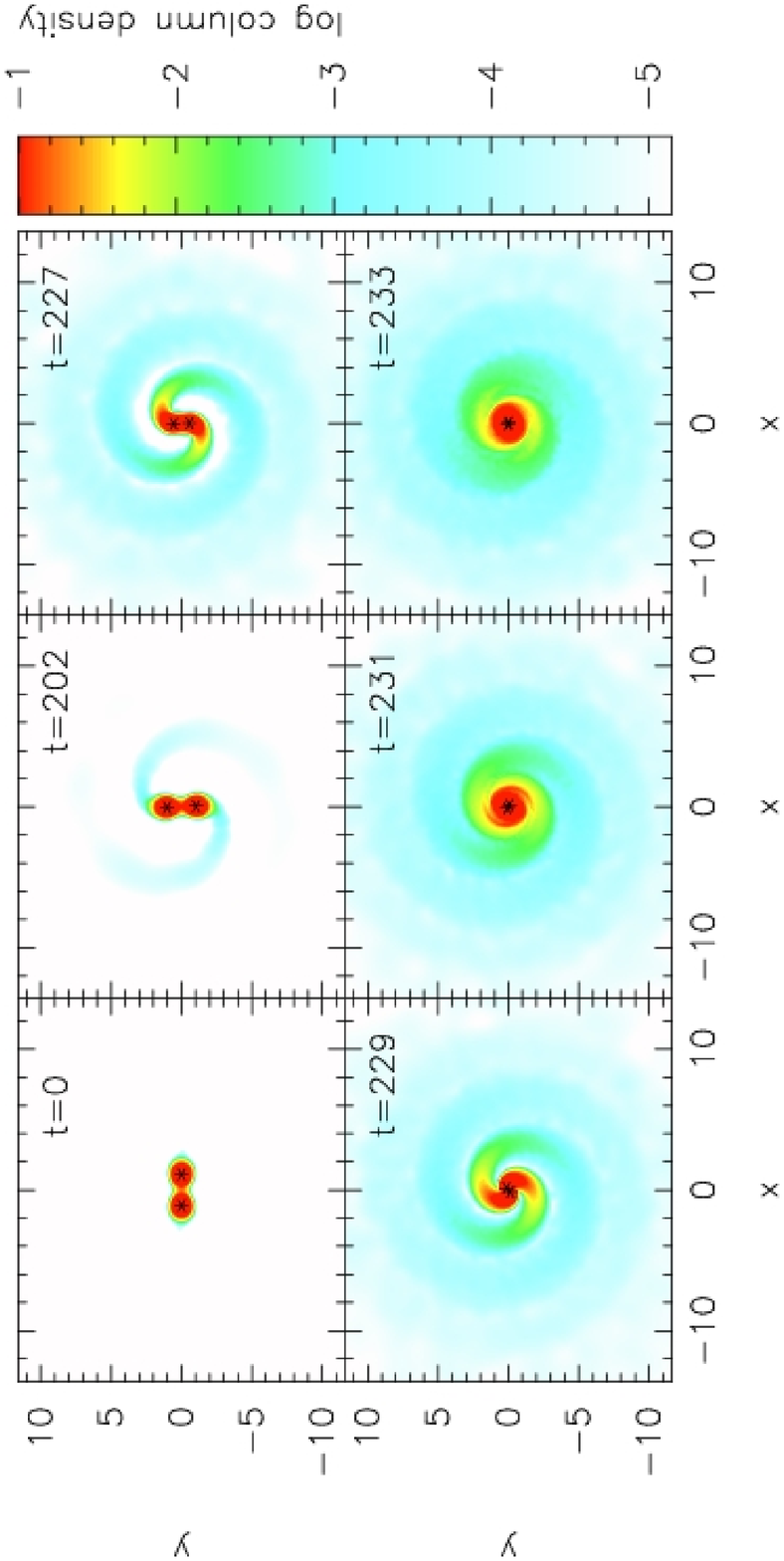}
\caption{Like Fig.~\ref{fs0.050_r2.54b}, but for the $m_c=0.2$
  dynamical calculation starting at $r=2.20$, just inside the Roche limit.
\label{fs0.200_r2.20}
}
\end{figure*}

\begin{figure}
\includegraphics[width=4in]{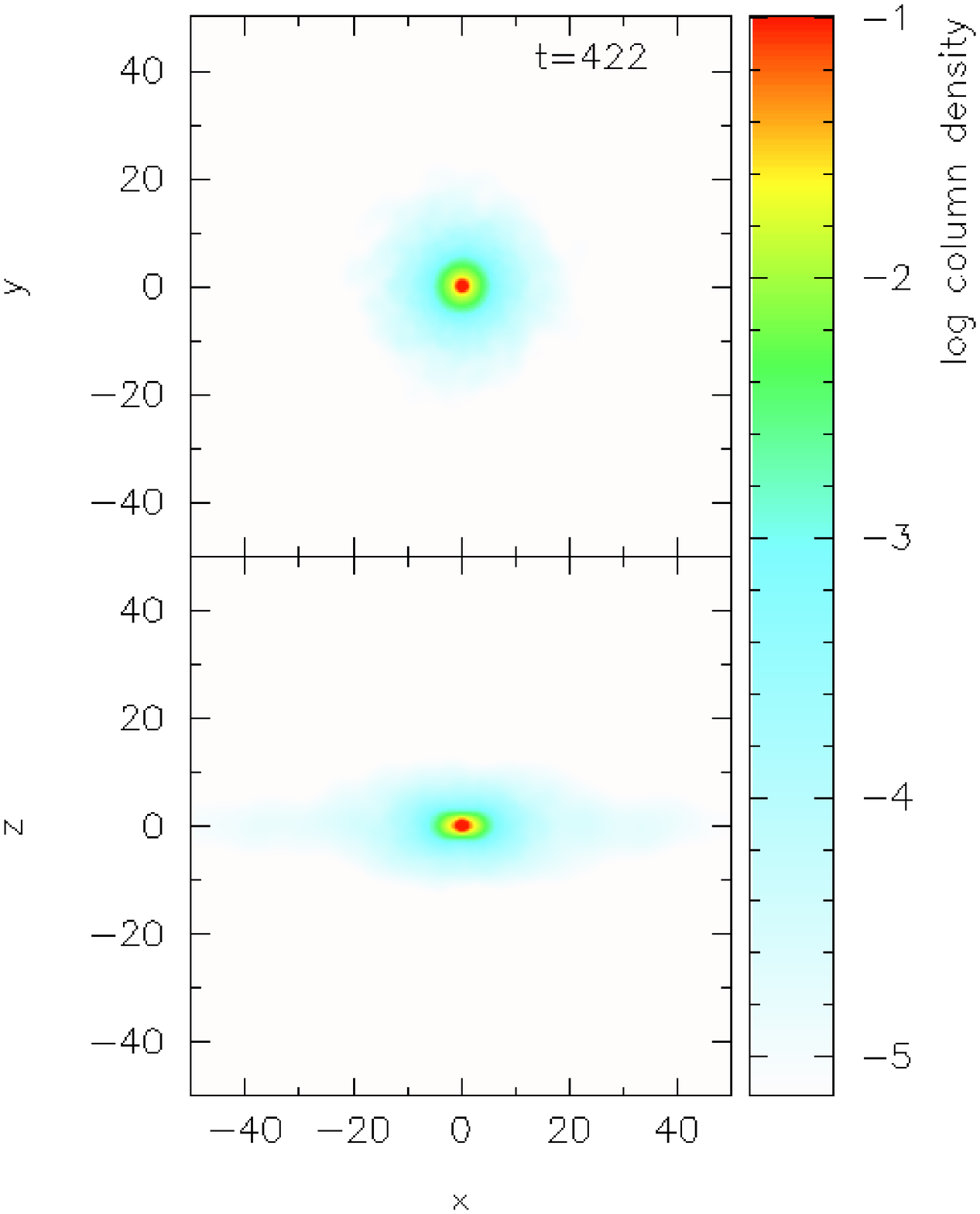}
\caption{The state of the simulation presented in
  Fig.~\ref{fs0.200_r2.20} at a late time ($t=422$), with the column density projected in both
  the $xy$ and $xz$ planes.
\label{fs0.200_r2.20_redo_longer_final_state}
}
\end{figure}

Figure~\ref{fs0.200_r2.20} presents both particle positions and column
density plots at six different times in the $m_c=0.2$, $r=2.20$
dynamical calculation, a case just inside the Roche limit. The
coordinate system used here rotates counterclockwise with a period of
14.22 time units, which equals the orbital period of this binary at
early times.  Gas is excreted almost immediately, with each parcel of
gas carrying a specific angular momentum essentially equal to that of
the outer Lagrangian points.  In contrast to mergers with $m_c\lesssim
0.15$, the excreted gas originates equally from both binary components
and flows past the outer Lagrangian points symmetrically.  As the
outer Lagrangian points are the outermost positions at which gas can
be in static equilibrium, they are also the positions of largest
possible specific angular momentum in rigidly rotating equilibrium
twin binaries.  Consequently, the mass loss necessarily decreases the
average angular momentum per unit mass of the gas remaining within the
outer Lagrangian surface, causing the binary components, along with
their cores, to inspiral:  
see the appendix of \citet{1976ApJ...209..829W} for a rigorous analysis of the
angular momentum budget during mass excretion.
As the components get closer, the
excretion rate increases, and in addition the resulting arms become
more tightly wound (compare the $t=202$ frame to later ones).  By
$t=233$, the central regions of the binary components have effectively
merged.  At later times, the merger product approaches a rapidly
rotating, axisymmetric configuration (see
Fig.~\ref{fs0.200_r2.20_redo_longer_final_state}).

\begin{figure}
\includegraphics[width=3.4in]{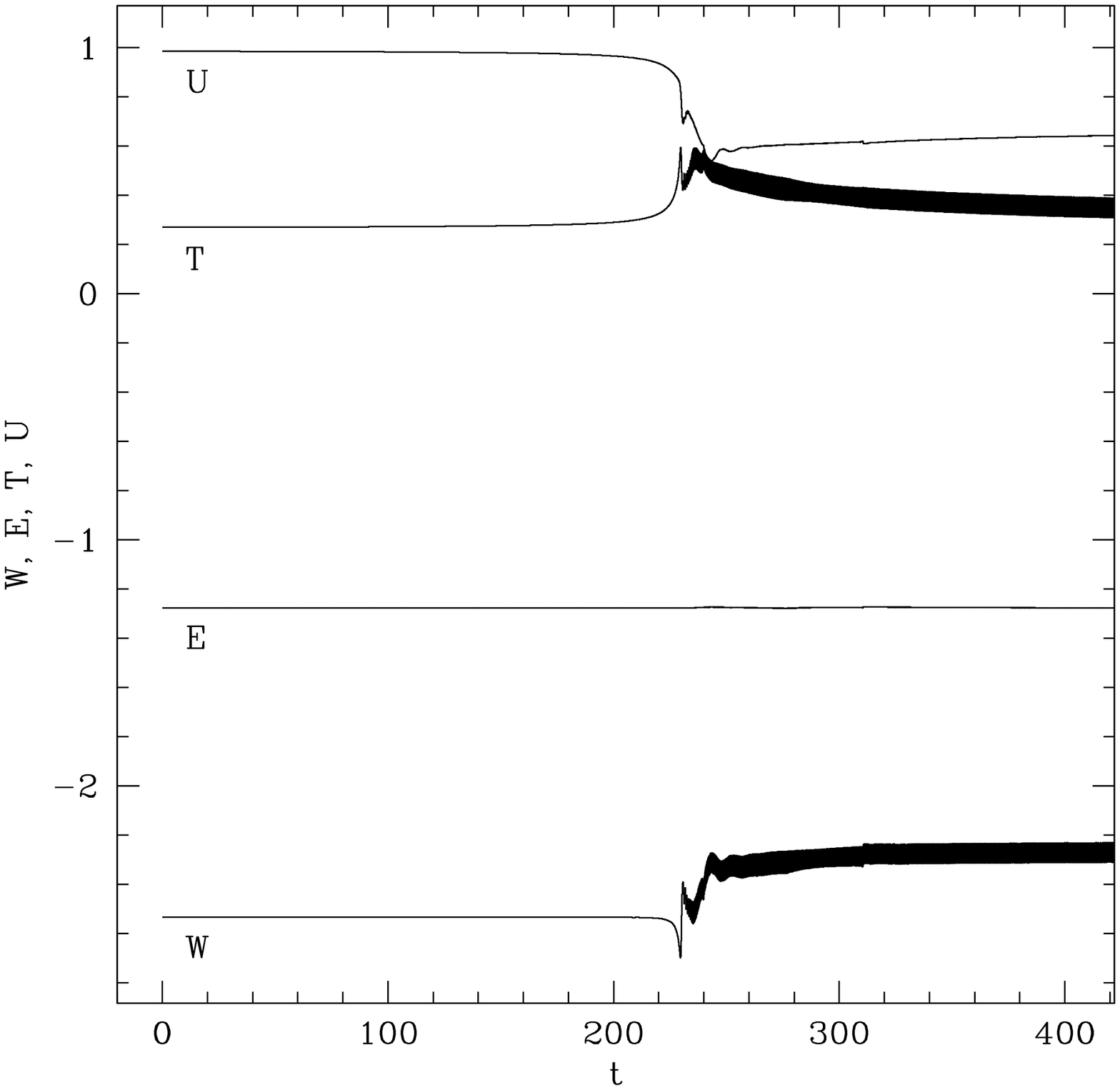}
\caption{Energies versus time $t$ for the simulation presented in
  Fig.~\ref{fs0.200_r2.20}.  From the bottom curve to the top one, we
  show gravitational energy $W$, total energy $E$, kinetic energy $T$,
  and internal energy $U$.
\label{fs0.200_r2.20_redo_longer_energies_10_334.59}
}
\end{figure}

Figure \ref{fs0.200_r2.20_redo_longer_energies_10_334.59} shows
energies versus time for the $m_c=0.2$, $r=2.20$ calculation.  The
kinetic energy $T$ gradually increases as the binary components
inspiral, until the cores approach closely at $t\approx 230$.  The
ensuing shocks cause the gas to expand, causing an overall decrease in
the internal energy $U$ and increase in the gravitational potential
energy $W$.  The rapid variations in $T$ and $W$ at late times are due
to the eccentric orbit of the central double core.  The total energy
$E$ is well conserved in this simulation, varying by only 0.4\% from
its minimum to maximum values over the entire time interval shown in
Figure \ref{fs0.200_r2.20_redo_longer_energies_10_334.59}.  Energy
conservation in other runs is typically at least this good and often
even much better.  In our simulations, most of the small
non-conservation in energy occurs at late times once the core
particles have entered a tight orbit.

\begin{figure}
\includegraphics[width=3.4in]{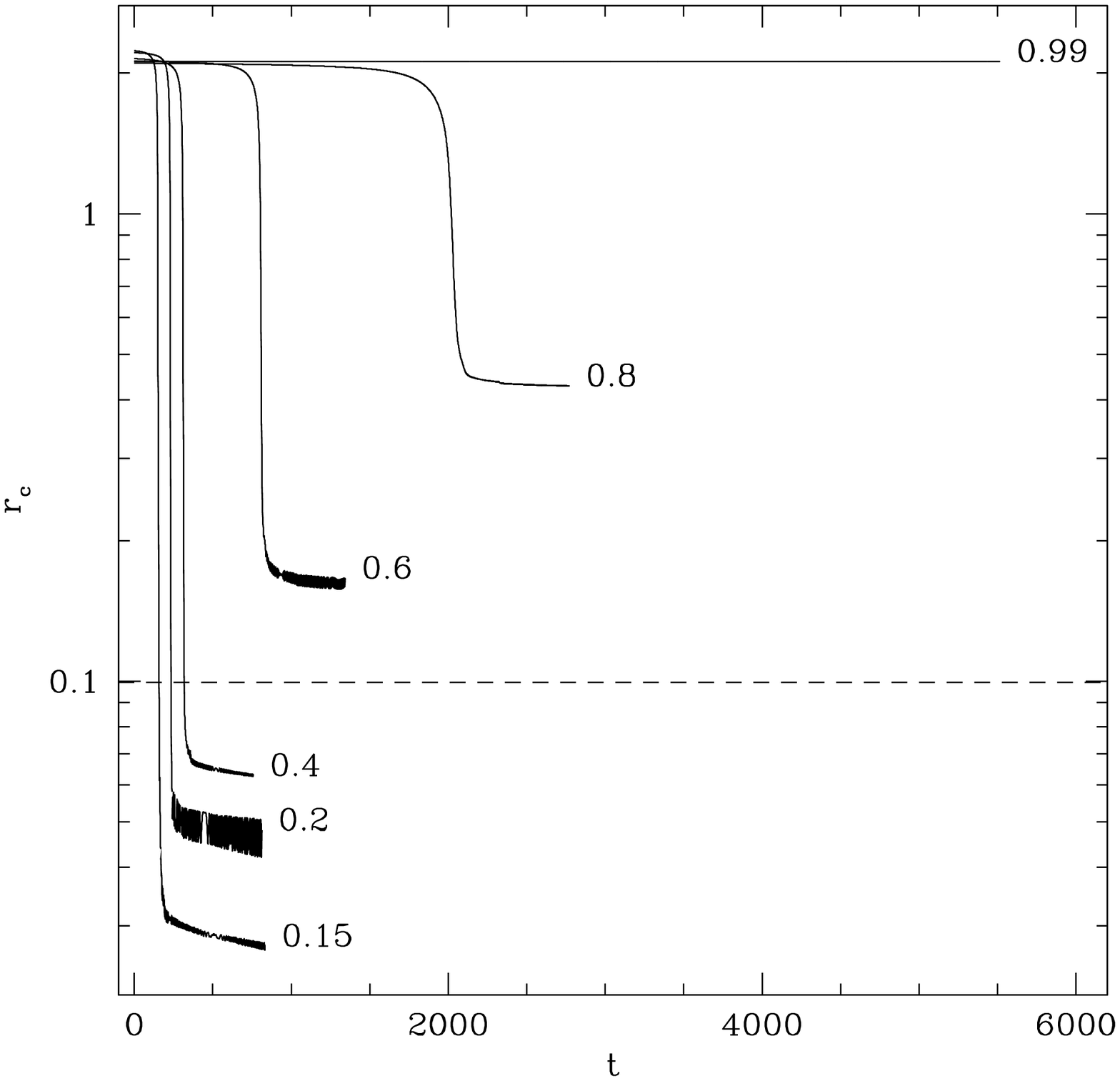}
\includegraphics[width=3.4in]{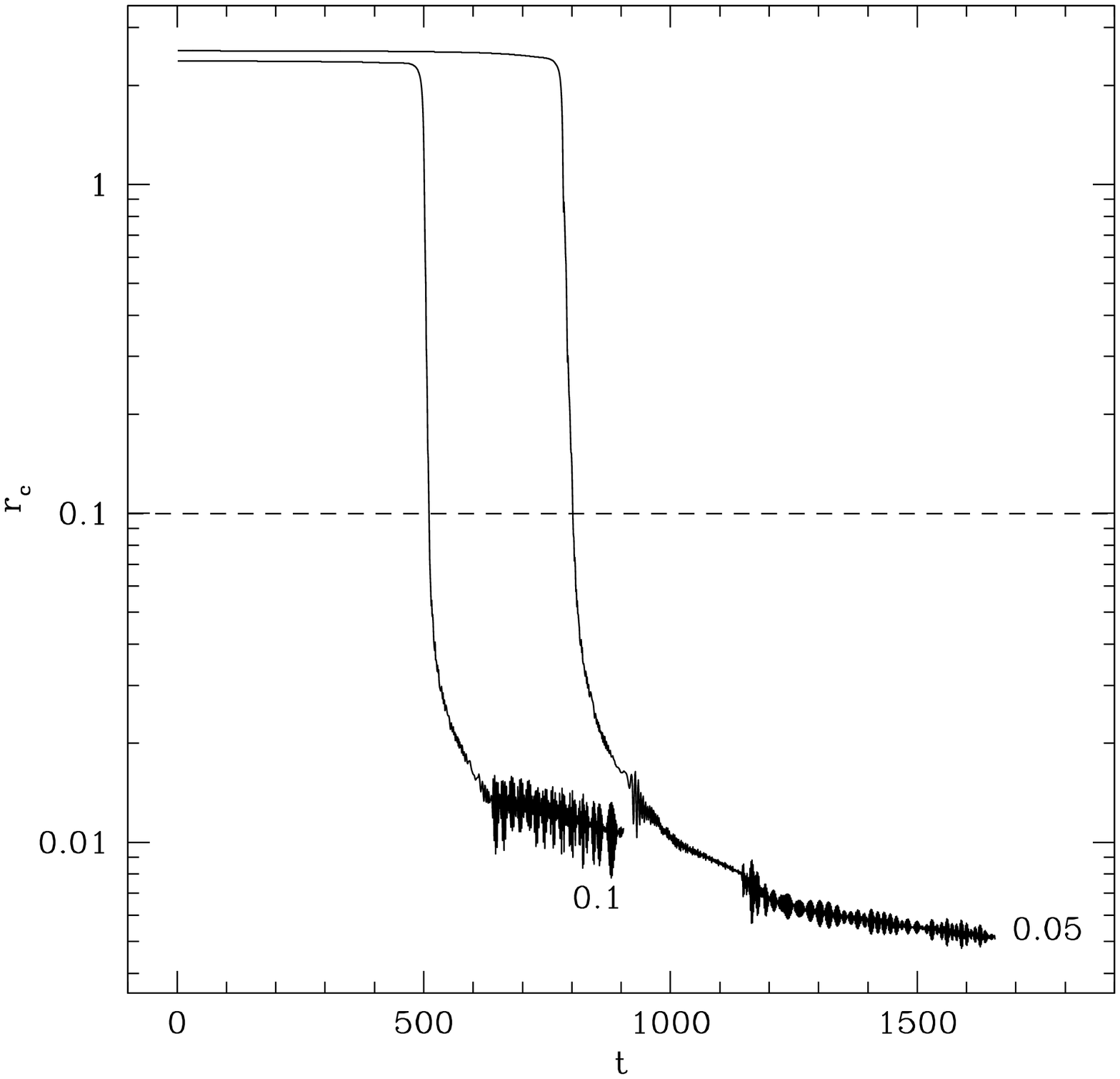}
\caption{The separation $r_c$ between the cores as a function of time $t$
  for binaries that begin just inside the Roche limit (top frame)
  and those that begin just inside the secular limit (bottom frame).
  Each curve is labeled by the factional core mass $m_c$.  The
  horizontal dashed line indicates the minimum separation for which the
  gravitational interaction of the two cores is unsoftened in our simulations.
\label{rc}
}
\end{figure}

In all our merger simulations of twin binaries with fractional core masses of
0.15 or larger, each star loses mass through an outer Lagrangian
point.  Most of this mass ultimately ends up in a circumbinary
envelope gravitationally bound to the central binary of cores.
The mass ejected varies from $\sim 0.5$\%
(for $m_c=0.15$) up to $\sim 7$\% (for $m_c=0.8$) of the total system mass.
A trend evident from the simulations is that more massive
cores ultimately remain more widely separated, after inspiraling, than
less massive cores.  The top frame of Figure \ref{rc} provides a
closer look at how the separation of the cores evolves in several
dynamical simulations that begin just inside the Roche limit.  For
very large fractional core masses ($m_c \gtrsim 0.9$), the gas is
simply not massive enough to affect significantly the dynamics of the
cores: although they inspiral, they stay separated at distances on the
order of the initial stellar radius.  For moderately large core masses ($0.5 \lesssim
m_c \lesssim 0.9$), the cores inspiral to a fraction of a stellar
radius, although the process halts at separations large enough still
to be resolved by our simulations.  For core masses in the range $0.15
\lesssim m_c \lesssim 0.5$, which corresponds to most red giants in
nature, the cores rapidly inspiral to separations less than 0.1.
Although our code has no mechanism for merging the cores, we do not
expect such an effect to be relevant here:
the size of a core relative to the stellar radius is
typically only $\sim 10^{-4}$ or less for a giant, so that a tremendous
amount of angular momentum would have to be removed from the double
core before they could merge.

The bottom frame of Figure \ref{rc} concerns fractional core masses
less than 0.15, namely those cases that reach the secular instability
limit before the Roche limit.
The cores in these merger simulations inspiral
to a separation $\lesssim 0.01$, considerably less than the spatial
resolution.  Whether or not the cores would merge in such
circumstances will likely depend on the details of the parent star
structure, with simulations that resolve the cores necessary to
address the issue fully.  We find that less than 0.5\% of the system
mass is ejected whenever the merger is triggered by mass transfer.

\begin{figure}
\includegraphics[width=3.4in]{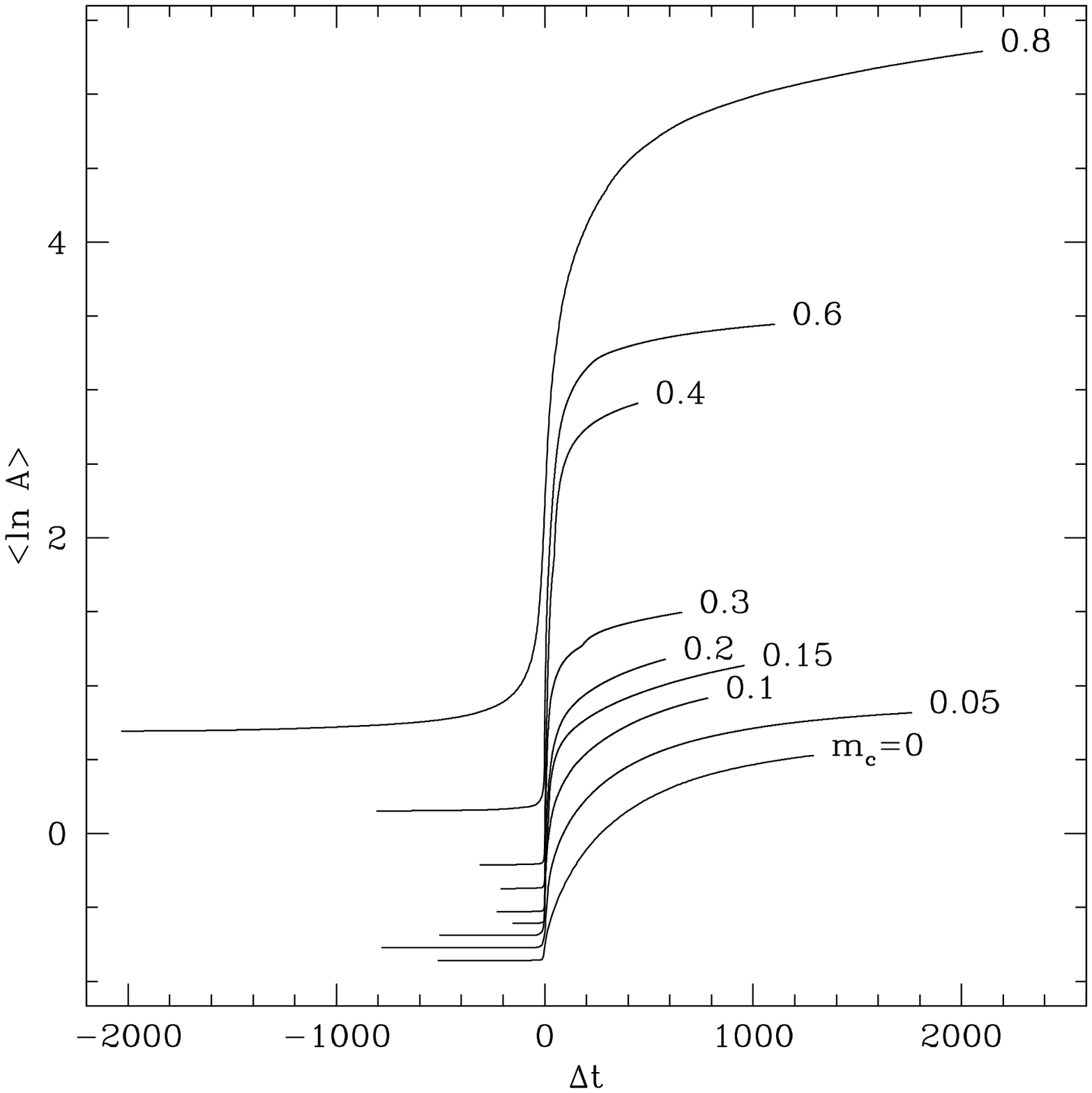}
\caption{A measure of entropy evolution due to shock heating during
  several mergers.  The natural logarithm of the entropic variable $A$
  is averaged by mass over the gas of the system and plotted versus a
  shifted time coordinate $\Delta t$, chosen such that the inflection
  point in each curve (when the entropy is increasing most rapidly)
  occurs at $\Delta t=0$. Each curve is labeled by the factional core
  mass $m_c$: as $m_c$ increases, so do both the initial and final
  values of $<\ln A>$.  For each core mass, we present our simulation
  of largest initial separation that results in a merger,
  corresponding in Fig.\ \ref{aa_20000} to the uppermost asterisk (for
  a given $m_c<0.15$) or filled square (for a given $m_c\ge 0.15$).
  Note that the initial value of the entropic variable $A$ equals the
  $K$ in equation (\ref{mrrelation}), as calculated from the
  appropriate $E_{\rm O}$ in Table \ref{parents}.  The entropic
  variable $A$ is in units of $GM^{1/3}R$.
\label{entropyall}
}
\end{figure}

Figure \ref{entropyall} helps to quantify the entropy evolution during
mergers by plotting the mass-average $<\ln A>$ over time for several
cases.  We note that, for our polytropic equation of state and gas of
uniform composition, the specific entropy of a parcel of gas is
proportional to $\ln A$ plus a constant.  From the curves of Figure
\ref{entropyall}, we see that the change in entropy per unit mass
tends to be larger, and develops on a longer timescale, for cases
involving larger core masses.  We also note that the entropy is
still gradually increasing even at the end of our simulations, due to
the influence of the central binary embedded within the system.

Like the critical separations for first contact and secular
instability, the Roche limit separation tends to decrease as the core
mass increases.  A fitted formula consistent with our 
dynamical integrations
to within $\sim 1$\% for any core mass $m_c$ is
\begin{equation}
r_{\rm Roche}\approx 2.11+0.25(1-m_c)^4.\label{rroche}
\end{equation}
Because the degree of contact $\eta$ varies
nearly linearly with separation $r$ (for two examples, see Fig.\
\ref{eta} of this paper and Fig.\ 2 of \citet{1995ApJ...438..887R}),
this formula, along with others from \S\ref{equilibrium_model_sequences}, allows us also to estimate the degree of contact
$\eta$ at the secular instability limit: $\eta_{\rm sec}\approx(r_{\rm
fc}-r_{\rm sec})/(r_{\rm fc}-r_{\rm Roche})$.

\section{Discussion and Future Work\label{future}}

\subsection{Summary of Main Results}

We have determined equilibrium sequences and performed dynamical
calculations of twin binaries, focusing primarily on configurations in
which the stars are in contact.  Our equilibrium sequences of
\S\ref{equilibrium_model_sequences} allow us to determine the binary
separation at first contact and at the innermost stable circular orbit
as a function of the fractional core mass $m_c$.
For $m_c\lesssim 0.15$, the
innermost stable orbit occurs at the secular instability limit (marked by
a minimum in energy and angular momentum along the equilibrium
sequence); for $m_c\gtrsim 0.15$, the
innermost stable orbit occurs at the Roche limit (defined as the
minimum separation for which an equilibrium configuration exists).
Our dynamical calculations of \S3.2 confirm these critical separations
and also reveal how the components inspiral once a binary passes the
innermost stable orbit.

\begin{figure}
\includegraphics[width=3.4in]{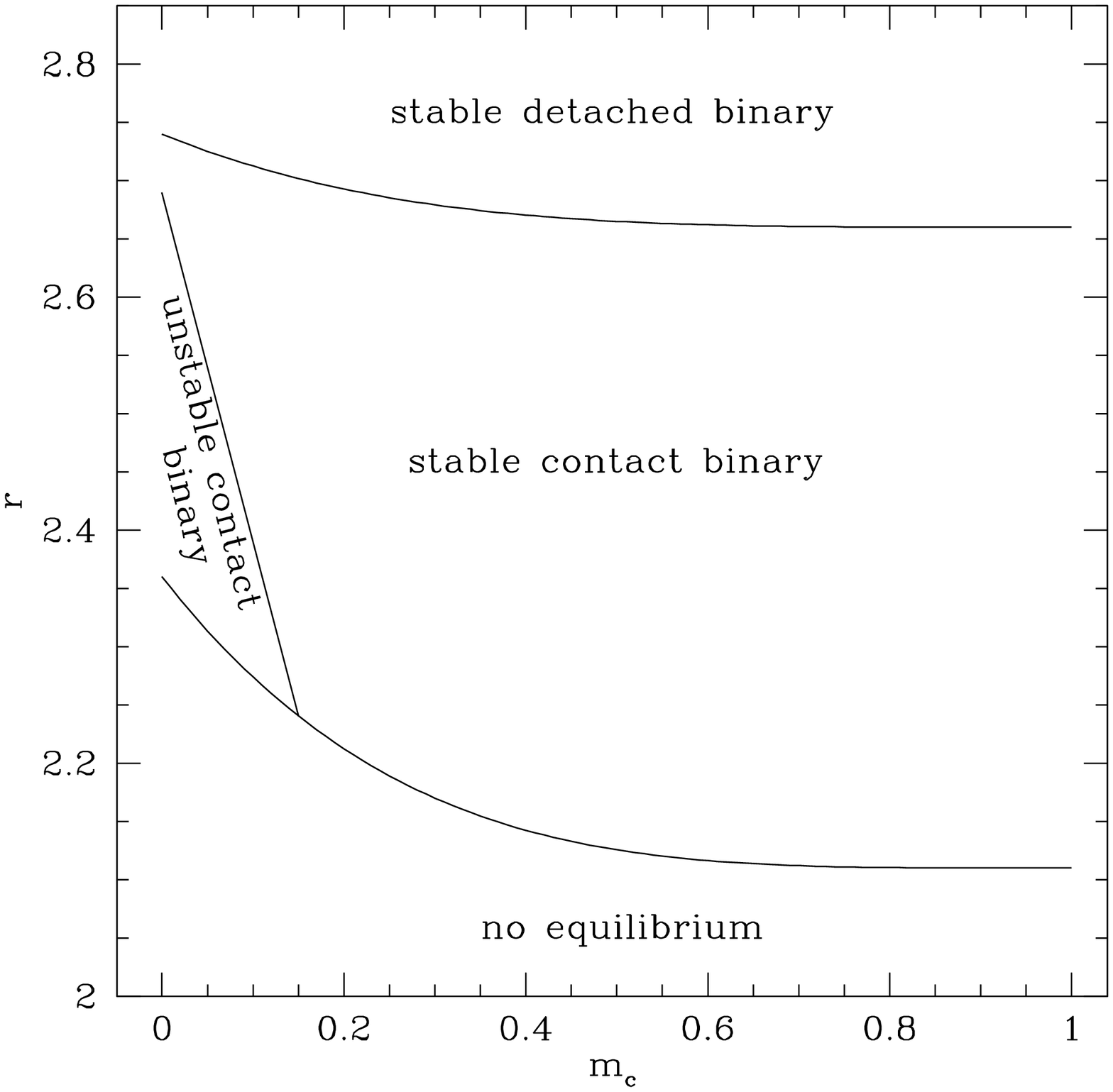}
\caption{The parameter space of twin binaries.
  Here and throughout the paper, the core mass $m_c$ and binary
  separation $r$ are given in units of the total mass and radius,
  respectively, of an isolated binary component. The top curve,
  separating detached and contact binaries, marks configurations of first contact.
  The middle curve, spanning $0\le m_c \lesssim 0.15$, is the secular
  instability limit that separates stable and unstable contact
  binaries: in this work, we find that most secularly unstable systems
  are also dynamically unstable to mass transfer across the inner
  Lagrangian point.  The bottom curve, separating contact binaries
  from configurations with no equilibrium, represents the Roche limit.
  Binaries that are unstable or that cannot exist in equilibrium 
  have their components inspiral and merge, a process we follow with
  dynamical calculations.
\label{aa_qualitative}
}
\end{figure}

Figure~\ref{aa_qualitative} summarizes graphically our most basic
results.  Recall that the separation $r$ on the vertical axis is
scaled to the unperturbed stellar radius $R$, while the core mass
$m_c$ on the horizontal mass is scaled to the stellar mass $M$.
Thus, as the components
in a twin binary expand and gradually increase their core masses due to stellar
evolution, the corresponding position in the parameter space of
Figure~\ref{aa_qualitative} will shift down and slightly to the
right.  When this position drops below the top curve, which marks
first contact, the binary enters
the stable contact phase.  When the position drops into either the
unstable or no equilibrium portions of parameter space, the
components merge.

The curves shown in Figure~\ref{aa_qualitative} are given by equations
(\ref{rfc}), (\ref{rsec}), and (\ref{rroche}); population modelers can
use these fitted formulae in treatments of twin binaries.  Consider,
for example, such a binary with a given orbital separation $a$.  The
stellar evolution of each component gives the time dependence of the
stellar radius $R$, the stellar mass $M$, and the core mass $M_c$.
The dimensionless separation $r=a/R$ and the fractional core mass
$m_c=M_c/M$ are thus known functions of time, and the evolutionary
track can be placed in the theoretical $r$ versus $m_c$ diagram
(Figure~\ref{aa_qualitative}) to determine the final fate of the
system.

We note that the volume of parameter space where real binaries would
ultimately end up in the stable contact region, without crossing the
instability limit or the Roche limit, is small.  Such a situation
would require a fine tuning of the initial semimajor axis $a$.  For
example, a star with an initial mass of $8M_\odot$ will expand to
$R\approx 370R_\odot$ and reach a fractional core mass of $m_c\approx
0.2$ as it ascends the asymptotic giant branch, according to
calculations by the TWIN stellar evolution code.  For this core mass,
$r_{\rm fc}\approx 2.7$ and $r_{\rm Roche}\approx 2.2$.  Thus, a twin
binary composed of such stars will remain detached if $a\gtrsim r_{\rm
  fc}R\approx 1000R_\odot$ and will ultimately surpass the Roche limit
if $a\lesssim r_{\rm Roche}R \approx 800R_\odot$.  Only if
$800R_\odot\lesssim a\lesssim 1000R_\odot$ will the binary reach the
contact phase without the cores also inspiraling to form a tight
binary.

The dynamic simulations of \S\ref{dynamical_integrations} always start
from a symmetric equilibrium configuration, with the binary components
being hydrostatic in the corotating frame.  In our simulations with
core masses of $m_c=0.125$ and less, we find a dynamical instability
to exist at or slightly inside the secular instability limit.  This
dynamical instability is a global instability of the equilibrium
state, triggered by small numerical noise and characterized by a
growing asymmetric mode.  Binaries that come inside this instability
limit first transfer mass gradually from one component to the other
and eventually coalesce quickly as mass is lost through the outer
Lagrangian points.  The cores are left in a tight binary surrounded by
a circumbinary envelope.

\subsection{Comparison with Other Works}

The merger of our small core mass twin binaries proceeds in a fashion
qualitatively similar to that of the Q1.3 model of
\cite{2006ApJ...643..381D}.  In that model, the binary is composed of
purely polytropic components ($m_c=0$) with mass ratio (donor to
accretor) $q=1.3$.  In both our simulations and theirs, the dynamical
instability manifests itself as a gradually developing mass transfer
flow, followed by excretion of gas through the outer Lagrangian points
and merger of the stellar envelopes.  One important difference,
however, is that the instability in our $m_c=0$ case does not develop
until the stars have reached a degree of contact $\eta\approx 0.17$,
whereas the instability is present in the Q1.3 model while the binary
is still semidetached.  This difference highlights the stabilizing
influence of the common envelope in twin binaries, even though the
instability still exists even for $q=1$.

We can directly compare our results for the limiting case $m_c=0$ with
those of \citet{1995ApJ...438..887R}.  The agreement in the first
contact, secular instability, and Roche limit separations is excellent
(better than 1\%).  The computational resources of the time, however,
limited \citet{1995ApJ...438..887R} to follow up to only $\sim 3$
orbits, so that they were unable to identify weak mass transfer
events, that is, events that develop gradually over many
dynamical timescales \citep{1972AcA....22...73P}.  As a result, they
determined the dynamical instability limit to be at $r\approx 2.45$,
well inside the secular instability limit.  In contrast, by following the dynamical
evolution of $m_c=0$ twin binaries for up to $\sim 100$ orbits, we
find that the dynamical instability limit actually coincides, or
nearly coincides, with the secular instability limit at $r\approx
2.67$.


\subsection{Relevance to Binary Neutron Stars}

We find that twin binaries
with $m_c\gtrsim 0.15$ exist stably at separations all the way down to the
Roche limit, where
mass is then excreted symmetrically through the outer Lagrangian
points.  This excretion carries away angular momentum and causes the
stars, along with their cores, to inspiral on a dynamical timescale.  For core masses
$m_c\lesssim 0.9$, the cores inspiral to a final separation that is a
fraction of the stellar radius.  Indeed for $m_c\lesssim0.5$, which
corresponds to most giant stars in nature, the final separation of the
cores is less than one-tenth the stellar radius.  Thus, we are left
with two cores in a tight binary surrounded by the combined gaseous
envelopes of the original binary, the precursor for double neutron
stars proposed in the \citet{1995ApJ...440..270B} scenario.

As the gravitational forces of the cores are softened at distances
less than $\sim 0.1$, our dynamic simulations that lead to cores in a
tight binary can provide only an upper limit on the separation at the
end of their inspiral.  
Our simulations of the $m_c=0.15$ case, for
example, place this upper limit at $\sim 0.03$ times the stellar
radius (see Fig.~\ref{rc}).  Thus, a binary composed of twin $M=10
M_\odot$, $R=200R_\odot$, $M_c=1.5M_\odot$ red giants would have their
cores spiral to a separation of less than $0.03R\approx6R_\odot$.  The
gradual transfer of energy to the circumbinary envelope could
easily decrease the separation of the cores by a factor
of $\sim 2$ further.  As gravitational radiation alone can bring two
$1.5 M_\odot$ point masses separated by up to $\sim5R_\odot$ to contact
in less than a Hubble time, such systems could have evolved to become
arbitrarily tight by the present time.  We therefore believe that
such double cores are indeed excellent candidates for
binary NS progenitors, as proposed in the Brown scenario.


\subsection{Relevance to Planetary Nebulae}

Given the relatively short timescale covered by hydrodynamic simulations such as ours,
 the circumbinary envelopes at the end
of our dynamical simulations are still optically thick.  Nevertheless,
 it is natural to
think of the final states of our merger calculations as a type of proto-PNe,
specifically, as
the immediate precursors of PNe with equal mass central binaries.
Future work could
use the end state of hydrodynamic calculations as initial
conditions in models of PN evolution, with particular
attention paid to any transition to the optically thin regime (revealing the central binary) and to
the morphology of the gas distribution.

To examine whether the dynamical calculations of this paper yield final states that
are consistent with the observations of the $q\approx 1$ double degenerate binaries, we consider a twin binary with $3M_\odot$ components.
We use the TWIN stellar evolution code to evolve each star in isolation.  The calculation shows that
the star expands to
only $R\approx 30R_\odot$ as it ascends the giant branch
and to nearly $500 R_\odot$ on the asymptotic giant branch.
Therefore, for any reasonable distribution
of initial orbital separations, twin binaries made of such stars would
much more often come into contact while on the asymptotic giant branch than when on the red giant
branch, consistent with the fact that most or all of the well-observed double degenerate stars in PNe seem to have masses too large to be He white dwarfs.
%
%

For the sake of discussion, consider such a twin binary that reaches
the Roche limit when $R=200R_\odot$.  According to the stellar
evolution calculation, an isolated star at that time would have a
carbon-oxygen core of mass $M_c=0.57M_\odot$ and, accounting for stellar winds, a total
mass $M=2.8M_\odot$.  At this fractional core mass
$m_c=0.57/2.8=0.20$, the dimensionless Roche separation (from
the results of this paper) is $r_{\rm Roche}\approx 2.2$, and thus the
initial semimajor axis of this binary would have been $a=r_{\rm
  Roche}R\approx 440 R_\odot$.  From the $m_c=0.2$ curve in the top
frame of Figure \ref{rc}, we see we can place an upper limit of $\sim
0.05R=10R_\odot$ on the final separation of the two 
cores, which, from Kepler's third law, corresponds to an upper limit
on the orbital period of 3 days.  If instead $R=100R_\odot$ at the time of merger, then a similar calculation gives a core mass $M_c=0.53M_\odot$ and an upper limit on the orbital period of 0.9 days.

For comparison, the central binaries in NGC 6026, Abell 41, and Hen 2-428 have comparable masses to these core masses
and orbital periods in roughly the 0.2 to 0.5 day range.  We conclude that
the results of our dynamical simulations, when applied to intermediate mass twin binaries, are consistent with
observed characteristics
of double degenerate central binaries in PNe.
In future simulations, a reduced gravitational softening between the two cores could allow for an even more precise determination of their final orbital separation.  In this way, observed orbital parameters of degenerate binaries in PNe could be more readily connected to the properties of the parent stars from which they may have originated.




\subsection{Additional Comparisons and Future Work}

For core masses $m_c\gtrsim 0.15$, we find that a twin binary can
exist stably in deep contact, at separations all the way down to the
Roche limit.  In contrast, the semi-analytic condensed polytrope models of
\citet{1987ApJ...318..794H} predict instead that twin binaries will
experience sustained mass transfer once the components come in
contact, provided only that $m_c<0.458$ (a range that includes the
vast majority of giant stars).  The primary oversimplification in the
semi-analytic treatment appears to be the approximation that mass
outside of the Roche lobe cannot help to contain the star within it.
Our numerical calculations, however, model the common envelope that
exists outside of the Roche lobe and that acts to suppress mass
transfer.  In addition, our fully three-dimensional calculations
remove the point mass and spherical structure approximations implicit
to the semi-analytic method.

The models of \citet{1987ApJ...318..794H} seem best suited to
semidetached binaries, where there is no common envelope to complicate
the dynamics of the mass flow.  A comparison of such cases
with our results is not possible, as our work is limited to binaries
with identical components.  Natural future work would include
relaxing this constraint so that binaries with mass ratio $q\ne 1$ can be
studied and compared with semi-analytic models.  Of particular
interest to the binary neutron star problem would be cases in which
the mass ratio deviated from one by only a few percent or less.

The modeling of giants as $\Gamma=5/3$ condensed polytropes is a
common simplifying approximation, one that here allows our results to
be scaled to binaries of any mass and length scales.  We note,
however, that radiation pressure can be the dominant contributor to
the equation of state at some ages and at some locations within the
envelopes of massive giants ($M\gtrsim 14M_\odot$, according to
calculations with the TWIN stellar evolution code). For such cases, our
treatment of the envelope as a constant entropy, $\Gamma=5/3$ gas can
be legitimately questioned.  While the effects of employing more
realistic stellar models would be worthwhile to study in future work,
we do not expect our results to change qualitatively.  Regardless of
the equation of state, gas that flows past the outer Lagrangian points
will still necessarily carry away a specific angular momentum larger
than the system average, forcing the remaining gas to configurations
of smaller angular momentum per unit mass.  We conclude that the
inspiral of cores should be a common outcome whenever a real twin binary
exceeds the Roche limit.

\acknowledgments

We thank Evghenii Gaburov, Zachary Proulx, Adam Simbeck, Eric
Theriault, and the anonymous referee for helpful input.
This material is based upon work
supported by the National Science Foundation under grant no.\ 0703545
and has made use of both the
SPLASH visualization software \citep{2007PASA...24..159P}
and NASA's Astrophysics Data System.
F.A.R.\ acknowledges support from NSF grant PHY--0855592.
V.K.\ acknowledges support from NSF grant PHY--0969820.


\begin{thebibliography}{}

\bibitem[Athanassoula et al.(2000)]{2000MNRAS.314..475A}
Athanassoula, E., 
Fady, E., Lambert, J.~C., \& Bosma, A.\ 2000, \mnras, 314, 475 

\bibitem[Belczynski et al.(2002)]{2002ApJ...572..407B} Belczynski, K., 
Kalogera, V., \& Bulik, T.\ 2002, \apj, 572, 407 

\bibitem[Burgay et al.(2003)]{2003Natur.426..531B} Burgay, M., et al.\ 
2003, \nat, 426, 531 

\bibitem[Bethe \& Brown(1998)]{1998ApJ...506..780B}
Bethe, H.~A., \& Brown, 
G.~E.\ 1998, \apj, 506, 780 

\bibitem[\protect\citeauthoryear{Bethe, Brown, 
\& Lee}{2007}]{2007PhR...442....5B} Bethe H.~A., Brown G.~E., Lee C.-H., 2007, PhR, 442, 5 

\bibitem[Bhattacharya \& van den Heuvel(1991)]{1991PhR...203....1B} 
Bhattacharya, D., \& van den Heuvel, E.~P.~J.\ 1991, \physrep, 203, 1 

\bibitem[Brown(1995)]{1995ApJ...440..270B} Brown, G.~E.\ 1995, \apj, 440, 
270

\bibitem[Bruch et 
al.(2001)]{2001A&A...377..898B} Bruch, A., Vaz, L.~P.~R., \& Diaz, M.~P.\ 2001, \aap, 377, 898 B

\bibitem[Chandrasekhar(1939)]{1939isss.book.....C} Chandrasekhar, S.\ 1939, 
An introduction to the study of stellar structure
(Chicago: The University of Chicago press)

\bibitem[Chandrasekhar(1969)]{1969efe..book.....C} Chandrasekhar, S.\ 1969, 
Ellipsoidal Figures of Equilibrium, New Haven: Yale University Press, 1969 (Revised Dover edition 1987)

\bibitem[Chandrasekhar(1975)]{1975ApJ...202..809C} Chandrasekhar, S.\ 1975, 
\apj, 202, 809 

\bibitem[Chevalier(1993)]{1993ApJ...411L..33C} Chevalier, R.~A.\ 1993, 
\apjl, 411, L33 

\bibitem[Counselman(1973)]{1973ApJ...180..307C} Counselman, C.~C., III 
1973, \apj, 180, 307 

\bibitem[Dan et al.(2009)]{2009JPhCS.172a2034D} Dan, M., Rosswog, S., 
\& Br{\"u}ggen, M.\ 2009, Journal of Physics Conference Series, 172, 012034 

\bibitem[De Marco(2009)]{2009PASP..121..316D} De Marco, O.\ 2009, \pasp, 
121, 316 

\bibitem[Dehnen(2001)]{2001MNRAS.324..273D}
Dehnen, W.\ 2001, \mnras, 324, 273

\bibitem[Dewi \& van den Heuvel(2004)]{2004MNRAS.349..169D} Dewi, J.~D.~M., 
\& van den Heuvel, E.~P.~J.\ 2004, \mnras, 349, 169 

\bibitem[Dewi et al.(2006)]{2006MNRAS.368.1742D}
Dewi, J.~D.~M., 
Podsiadlowski, P., \& Sena, A.\ 2006, \mnras, 368, 1742 

\bibitem[D'Souza et al.(2006)]{2006ApJ...643..381D} D'Souza, M.~C.~R., 
Motl, P.~M., Tohline, J.~E., \& Frank, J.\ 2006, \apj, 643, 381 

\bibitem[\protect\citeauthoryear{Eggleton}{1971}]{1971MNRAS.151..351E} 
Eggleton P.~P., 1971, MNRAS, 151, 351

\bibitem[\protect\citeauthoryear{Eggleton}{1983}]{1983ApJ...268..368E} 
Eggleton P.~P., 1983, ApJ, 268, 368 

\bibitem[Frank(2008)]{2008NewAR..51..878F} Frank, J.\ 2008, New Astronomy 
Review, 51, 878 

\bibitem[Fryer et al.(1996)]{1996ApJ...460..801F} Fryer, C.~L., Benz, W., 
\& Herant, M.\ 1996, \apj, 460, 801

\bibitem[Gaburov et al.(2010)]{2009arXiv0904.0997G} Gaburov, E., Lombardi, 
J., \& Portegies Zwart, S.\ 2010, MNRAS, 402, 105

\bibitem[Geier et 
al.(2011)]{2011A&A...528L..16G} Geier, S., Napiwotzki, R., Heber, U., \& Nelemans, G.\ 2011, \aap, 528, L16 

\bibitem[\protect\citeauthoryear{Glebbeek, Pols, 
\& Hurley}{2008}]{2008A&A...488.1007G} Glebbeek E., Pols O.~R., Hurley J.~R., 2008, A\&A, 488, 1007 

\bibitem[Habets(1986)]{1986A&A...167...61H} Habets, G.~M.~H.~J.\ 1986, 
\aap, 167, 61 

\bibitem[Hachisu \& Eriguchi(1984)]{1984PASJ...36..239H} Hachisu, I., \& 
Eriguchi, Y.\ 1984, \pasj, 36, 239 


\bibitem[H{\" a}rm \& Schwarzschild(1955)]{1955ApJS....1..319H} H{\" a}rm, R., \& 
Schwarzschild, M.\ 1955, \apjs, 1, 319 

\bibitem[Hernquist \& Katz(1989)]{1989ApJS...70..419H}
Hernquist, L., \& 
Katz, N.\ 1989, \apjs, 70, 419

\bibitem[Hillwig(2011)]{2011apn5.confE.275H} Hillwig, T.~C.\ 2011, 
Asymmetric Planetary Nebulae 5 Conference

\bibitem[Hillwig et al.(2010)]{2010AJ....140..319H} Hillwig, T.~C., Bond, 
H.~E., Af{\c s}ar, M., \& De Marco, O.\ 2010, \aj, 140, 319

\bibitem[Hjellming \& Webbink(1987)]{1987ApJ...318..794H} Hjellming, M.~S., 
\& Webbink, R.~F.\ 1987, \apj, 318, 794

\bibitem[Hogeveen(1992a)]{1992Ap&SS.194..143H} Hogeveen, S.~J.\ 1992, \apss, 
194, 143 

\bibitem[Hogeveen(1992b)]{1992Ap&SS.195..359H} Hogeveen, S.~J.\ 1992, \apss, 
195, 359 

\bibitem[Hogeveen(1992c)]{1992Ap&SS.196..299H} Hogeveen, S.~J.\ 1992, \apss, 
196, 299 

\bibitem[Hulse \& Taylor(1975)]{1975ApJ...195L..51H} Hulse, R.~A., \& 
Taylor, J.~H.\ 1975, \apjl, 195, L51 

\bibitem[Hut(1980)]{1980A&A....92..167H} Hut, P.\ 1980, \aap, 92, 167 

\bibitem[Jacoby et al.(2006)]{2006ApJ...644L.113J} Jacoby, B.~A., Cameron, 
P.~B., Jenet, F.~A., Anderson, S.~B., Murty, R.~N., \& Kulkarni, S.~R.\ 
2006, \apjl, 644, L113 

\bibitem[Jones et al.(2010)]{2010MNRAS.408.2312J} Jones, D., et al.\ 2010, 
\mnras, 408, 2312

\bibitem[Kramer et al.(2006)]{2006Sci...314...97K} Kramer, M., et al.\ 
2006, Science, 314, 97 

\bibitem[Krumholz 
\& Thompson(2007)]{2007ApJ...661.1034K} Krumholz, M.~R., \& Thompson, T.~A.\ 2007, \apj, 661, 1034 

\bibitem[Lai et al.(1993a)]{1993ApJS...88..205L} Lai, D., Rasio, F.~A., \& 
Shapiro, S.~L.\ 1993, \apjs, 88, 205 


\bibitem[Lai et al.(1993b)]{1993ApJ...406L..63L} Lai, D., Rasio, F.~A., \& 
Shapiro, S.~L.\ 1993, \apjl, 406, L63 


\bibitem[Lai et al.(1994a)]{1994ApJ...420..811L} Lai, D., Rasio, F.~A., \& 
Shapiro, S.~L.\ 1994, \apj, 420, 811 


\bibitem[Lai et al.(1994b)]{1994ApJ...423..344L} Lai, D., Rasio, F.~A., \& 
Shapiro, S.~L.\ 1994, \apj, 423, 344 


\bibitem[Lai et al.(1994c)]{1994ApJ...437..742L} Lai, D., Rasio, F.~A., \& 
Shapiro, S.~L.\ 1994, \apj, 437, 742 

\bibitem[\protect\citeauthoryear{Lee, Park, 
\& Brown}{2007}]{2007ApJ...670..741L} Lee C.-H., Park H.-J., Brown G.~E., 2007, ApJ, 670, 741 

\bibitem[Levine et al.(1993)]{1993ApJ...410..328L} Levine, A., Rappaport, 
S., Deeter, J.~E., Boynton, P.~E., \& Nagase, F.\ 1993, \apj, 410, 328 


\bibitem[Lombardi et al.(1999)]{lsrs99}
Lombardi, J.\ C., Sills, A., Rasio, F.\ A., \& Shapiro, S.\ L.\ 1999,
 J.\ Comp.\ Phys., 152, 687


\bibitem[Lor{\'e}n-Aguilar et 
al.(2009)]{2009A&A...500.1193L} Lor{\'e}n-Aguilar, P., Isern, J., \& Garc{\'{\i}}a-Berro, E.\ 2009, \aap, 500, 1193 

\bibitem[Miszalski et 
al.(2009)]{2009A&A...505..249M} Miszalski, B., Acker, A., Parker, Q.~A., \& Moffat, A.~F.~J.\ 2009, \aap, 505, 249




\bibitem[Motl et al.(2002)]{2002ApJS..138..121M} Motl, P.~M., Tohline, 
J.~E., \& Frank, J.\ 2002, \apjs, 138, 121 

\bibitem[Nice et al.(1996)]{1996ApJ...466L..87N} Nice, D.~J., Sayer, R.~W., 
\& Taylor, J.~H.\ 1996, \apjl, 466, L87 

\bibitem[Osterbrock(1953)]{1953ApJ...118..529O}
Osterbrock, D.~E.\ 1953, 
\apj, 118, 529 

\bibitem[Ostriker \& Gunn(1969)]{1969ApJ...157.1395O}
Ostriker, J.~P., \& 
Gunn, J.~E.\ 1969, \apj, 157, 1395

\bibitem[Paczyi{\'n}ski 
\& Sienkiewicz(1972)]{1972AcA....22...73P} Paczyi{\'n}ski, B., \& Sienkiewicz, R.\ 1972, Acta Astronomica, 22, 73

\bibitem[Pinsonneault \& Stanek(2006)]{2006ApJ...639L..67P}
Pinsonneault, 
M.~H., \& Stanek, K.~Z.\ 2006, \apjl, 639, L67

\bibitem[Piran \& Shaviv(2005)]{2005PhRvL..94e1102P} Piran, T., \& Shaviv, 
N.~J.\ 2005, Physical Review Letters, 94, 051102 

\bibitem[Pols(1994)]{1994A&A...290..119P} Pols, O.~R.\ 1994, \aap, 290, 119 

\bibitem[\protect\citeauthoryear{Portegies Zwart et 
al.}{2009}]{2009NewA...14..369P} Portegies Zwart S., et al., 2009, NewA, 
14, 369

\bibitem[Price(2007)]{2007PASA...24..159P} Price, D.~J.\ 2007, Publ.\ Astron.\ Soc.\ Aust., 24, 
159 

\bibitem[Rasio(1991)]{ras91} Rasio, F.\ A.\ 1991, PhD Thesis, Cornell University

\bibitem[Rasio(1994)]{1994MmSAI..65...37R} Rasio, F.~A.\ 1994, Memorie 
della Societa Astronomica Italiana, 65, 37 

\bibitem[Rasio \& Lombardi(1999)]{ras99}
Rasio, F.\ A., \& Lombardi, J.\ C.\ 1999,
J.\ Comp.\ App.\ Math., 109, 213

\bibitem[Rasio \& Shapiro(1991)]{rs91}
Rasio, F.\ A., \& Shapiro, S.\ L.\ 1991, \apj, 377, 559

\bibitem[Rasio \& Shapiro(1992)]{1992ApJ...401..226R} Rasio, F.~A., \& 
Shapiro, S.~L.\ 1992, \apj, 401, 226 

\bibitem[Rasio \& Shapiro(1994)]{1994ApJ...432..242R} Rasio, F.~A., \& 
Shapiro, S.~L.\ 1994, \apj, 432, 242

\bibitem[Rasio \& Shapiro(1995)]{1995ApJ...438..887R} Rasio, F.~A., \& 
Shapiro, S.~L.\ 1995, \apj, 438, 887 

\bibitem[\protect\citeauthoryear{Rosswog}{2009}]{2009NewAR..53...78R} 
Rosswog S., 2009, NewAR, 53, 78

\bibitem[Rucinski(1992)]{ruc92}Rucinski, S.~M.\ 1992, in The Realm of Interacting Binary Stars,
eds.\ J.~Sahade et al.\ (Dordrecht: Kluwer), 177, 111


\bibitem[Santander-Garcia et al.(2011)]{2011apn5.confE.259S} 
Santander-Garcia, M., Rodr{\'{\i}}guez-Gil, P., Jones, D., Corradi, 
R.~L.~M., Miszalski, B., Pyrzas, S., 
\& Rubio-D{\'{\i}}ez, M.~M.\ 2011, Asymmetric Planetary Nebulae 5 Conference

\bibitem[\protect\citeauthoryear{Sch{\"o}nberg 
\& Chandrasekhar}{1942}]{1942ApJ....96..161S} Sch{\"o}nberg M., Chandrasekhar S., 1942, ApJ, 96, 161

\bibitem[Shimanskii et al.(2008)]{2008ARep...52..479S} Shimanskii, V.~V., 
Borisov, N.~V., Sakhibullin, N.~A., 
\& Sheveleva, D.~V.\ 2008, Astronomy Reports, 52, 479 


\bibitem[Stairs et al.(2002)]{2002ApJ...581..501S} Stairs, I.~H., Thrsett, 
S.~E., Taylor, J.~H., \& Wolszczan, A.\ 2002, \apj, 581, 501 

\bibitem[Stairs et al.(2006)]{2006MNRAS.373L..50S} Stairs, I.~H., Thorsett, 
S.~E., Dewey, R.~J., Kramer, M., \& McPhee, C.~A.\ 2006, \mnras, 373, L50 

\bibitem[Tassoul(1975)]{1975ApJ...202..803T} Tassoul, M.\ 1975, \apj, 202, 
803 

\bibitem[Thorsett \& Chakrabarty(1999)]{1999ApJ...512..288T} Thorsett, 
S.~E., \& Chakrabarty, D.\ 1999, \apj, 512, 288 

\bibitem[\protect\citeauthoryear{Wang, Lai, 
\& Han}{2006}]{2006ApJ...639.1007W} Wang C., Lai D., Han J.~L., 2006, ApJ, 639, 1007

\bibitem[Webbink(1976)]{1976ApJ...209..829W} Webbink, R.~F.\ 1976, \apj, 
209, 829

\bibitem[Webbink(2006)]{2006JAVSO..35..124W} Webbink, R.~F.\ 2006, Journal 
of the American Association of Variable Star Observers (JAAVSO), 35, 124 

\bibitem[Weisberg \& Taylor(2005)]{2005ASPC..328...25W} Weisberg, J.~M., \& 
Taylor, J.~H.\ 2005, ASP Conf.~Ser.~328: Binary Radio Pulsars, 328, 25 

\bibitem[Willems \& Kalogera(2004)]{2004ApJ...603L.101W} Willems, B., \& 
Kalogera, V.\ 2004, \apjl, 603, L101 

\bibitem[Willems et al.(2004)]{2004ApJ...616..414W} Willems, B., Kalogera, 
V., \& Henninger, M.\ 2004, \apj, 616, 414 

\bibitem[Wong et al.(2010)]{2010ApJ...721.1689W} Wong, T.-W., Willems, B., 
\& Kalogera, V.\ 2010, \apj, 721, 1689 


\end{thebibliography}
\end{document}